\pdfoutput=1
\newcommand*{\ATLASLATEXPATH}{latex/}
\documentclass[cernpreprint, texlive=2016, UKenglish]{\ATLASLATEXPATH atlasdoc}
\usepackage{\ATLASLATEXPATH atlaspackage}
\usepackage{\ATLASLATEXPATH atlasbiblatex}

\usepackage{\ATLASLATEXPATH atlasphysics}
\graphicspath{{logos/}{figures/}}

\usepackage{STDM-2018-03-PAPER-defs}

\AtlasTitle{Measurement of \wz production cross sections and gauge boson polarisation in $pp$ collisions at $\rts = 13~\tev$ with the ATLAS detector}

\AtlasAbstract{%
This paper presents measurements of \wz production cross sections in $pp$ collisions at a centre-of-mass energy of $13$~\TeV. 
The data were  collected in $2015$ and $2016$ by the ATLAS experiment  at the Large Hadron Collider, and correspond  to an integrated luminosity of {\mbox {$36.1$~fb$^{-1}$}}.
The \wz\ candidate events are reconstructed using leptonic decay modes of the gauge bosons into  electrons and muons.
The measured inclusive cross section in the detector fiducial region for a single leptonic decay mode is
$\sigma_{W^\pm Z \rightarrow \ell^{'} \nu \ell \ell}^{\textrm{fid.}} =  63.7  \, \pm~1.0~\textrm{(stat.)} \, \pm~2.3~\textrm{(syst.)} \, \pm~1.4~\textrm{(lumi.)}$~fb,
reproduced by the next-to-next-to-leading-order Standard Model prediction of $61.5^{+1.4}_{-1.3}$~fb.
Cross sections for $W^+Z$ and $W^-Z$ production and their ratio are presented as well as differential cross sections for several kinematic observables.
An analysis of angular distributions of leptons from decays of $W$ and $Z$ bosons is performed for the first time in pair-produced events in hadronic
collisions, and integrated helicity fractions in the detector fiducial region are measured for the $W$ and $Z$ bosons separately.
Of particular interest, the longitudinal helicity fraction of pair-produced vector bosons is also measured. 
}

\author{The ATLAS Collaboration}

\AtlasRefCode{STDM-2018-03}

\PreprintIdNumber{CERN-EP-2018-327}

\AtlasJournalRef{\EPJC 79 (2019) 535}
\AtlasDOI{10.1140/epjc/s10052-019-7027-6}

\hypersetup{pdftitle={ATLAS document},pdfauthor={The ATLAS Collaboration}}

\begin{document}

\maketitle

%\tableofcontents

%-------------------------------------------------------------------------------
\section{Introduction}
\label{sec:intro}
%-------------------------------------------------------------------------------

The study of  \wz diboson production is an important test of the Standard Model (SM) for its sensitivity to gauge
boson self-interactions, related to the non-Abelian structure of the electroweak interaction.
It provides the means to directly probe the triple gauge boson couplings (TGC), in particular the $WWZ$ gauge coupling.
Improved constraints from precise measurements can potentially probe scales of
new physics in the multi-\TeV\ range and provide a way to look for signals of new physics in a
model-independent way.
Previous measurements have concentrated on the inclusive and differential production cross sections.
In addition to more precise measurements of these cross sections that include new data, this paper presents measurements of the three helicity fractions of the $W$ and $Z$ bosons. 
The existence of the third state, the longitudinally polarised state, is a consequence of the non-vanishing mass of the bosons generated by the electroweak symmetry breaking mechanism. 
The measurement of the polarisation in diboson production therefore tests both the SM innermost gauge symmetry structure, through the existence of the triple gauge coupling, and the particular way this symmetry is spontaneously broken, via the longitudinal helicity state.
Angular observables can be used to look for new interactions that can lead to different polarisation behaviour than predicted by the SM, to which the \wz final state would be particularly sensitive~\cite{Azatov:2017kzw,Panico:2017frx}.
Precise calculations, at the next-to-leading order 
(NLO) in QCD, of SM polarisation observables in \wz production as well as electroweak corrections have recently appeared~\cite{Baglio:2018rcu}.
Polarisation measurements for each charge of the $W$ boson might be helpful in the investigation of $CP$ violation effects in the interaction between gauge bosons~\cite{Gounaris:1991ce,Kumar:2008ng}.
In the longer term, measuring the scattering of longitudinally polarised vector bosons will be a fundamental test of electroweak symmetry breaking~\cite{Ballestrero:2017bxn}.

Measurements of the \wz production cross section in proton--antiproton collisions
at a centre-of-mass energy of $\sqrt{s} = 1.96$~\TeV{} were published by
the CDF and D\O~collaborations~\cite{CDF:2012WZ,D0:2012WZ} using integrated luminosities of 
$7.1$~fb$^{-1}$ and $8.6$~fb$^{-1}$, respectively.
At the Large Hadron Collider (LHC), the most precise measurement of \wz production was reported by the ATLAS Collaboration~\cite{ATLASWZ_8TeV} using $20.1$~fb$^{-1}$ of data collected at a centre-of-mass energy of $8$~\TeV{}.
Measurements of \wz production at $\sqrt{s}=13$~TeV were reported by the ATLAS~\cite{Aaboud:2016yus} and CMS~\cite{Sirunyan:2019bez} 
collaborations using integrated luminosities of $3.2$~fb$^{-1}$ and $35.9$~fb$^{-1}$, respectively.
Other \wz measurements in $pp$ collisions, at centre-of-mass energies
of $7$~\TeV\ and $8$~\TeV, were reported previously by ATLAS 
and CMS~\cite{Aad:2012twa,Khachatryan:2016poo}.

At hadron colliders, the polarisation of the $W$ boson was previously measured in the decay of the top quark 
by the CDF and D\O~\cite{Aaltonen:2012rz,Aaltonen:2010ha,Abazov:2010jn} collaborations and the ATLAS~\cite{Aad:2012ky} and 
CMS~\cite{Khachatryan:2016fky} collaborations, as well as in association with jets by ATLAS~\cite{ATLAS:2012au} and CMS~\cite{Chatrchyan:2011ig}. 
Polarisation and several other related angular coefficient measurements of a singly produced $Z$ boson were published by the CDF~\cite{Aaltonen:2011nr}, CMS~\cite{Khachatryan:2015paa} and ATLAS~\cite{Aad:2016izn} collaborations. 
The polarisation of $W$ bosons was also measured in $ep$ collisions by the H1 Collaboration~\cite{Aaron:2009wp}.
Finally, for dibosons,  first measurements of the $W$ polarisation were performed by LEP experiments in $W$ pair production in $e^+e^-$ collisions~\cite{Achard:2002bv,Abbiendi:2003wv} and were used to set limits on anomalous triple gauge couplings (aTGC) in Ref.~\cite{Abdallah:2008sf}.

This paper presents results obtained using proton--proton ($pp$) collisions  recorded by the ATLAS detector at a
centre-of-mass  energy of $\sqrt{s}=13$~\TeV\ in $2015$ and $2016$, corresponding to an integrated luminosity of $36.1~\ifb$. 
The $W$ and $Z$ bosons are reconstructed using their decay modes into electrons or muons.
The production cross section is measured within a fiducial phase space both inclusively and differentially as a function of several individual variables related to the kinematics of the \wz\ system and to the jet activity in the event. 
The reported measurements are compared with the SM cross-section predictions at NLO in QCD~\cite{Ohnemus:1991,Frixione:1992} 
and with the recent calculations at next-to-next-to-leading order (NNLO) in QCD~\cite{Grazzini:2016swo, Grazzini:2017ckn}. 
The ratio of the $W^+Z$ cross section to the $W^-Z$ cross section, which is sensitive to the parton distribution functions (PDF) is also measured.
Finally, an analysis of angular distributions of leptons from decays of $W$ and $Z$ bosons is performed and integrated helicity fractions in the detector fiducial region are measured for the $W$ and $Z$ bosons separately.

%-------------------------------------------------------------------------------
\section{ATLAS detector}
\label{sec:detector}
%-------------------------------------------------------------------------------

The ATLAS detector~\cite{PERF-2007-01,IBL,Abbott:2018ikt}  is a multipurpose particle detector with a cylindrical 
geometry\footnote{ATLAS uses a right-handed coordinate system with its origin at the nominal 
interaction point (IP) in the centre of the detector and the $z$-axis along the beam direction. 
The $x$-axis points from the IP to the centre of the LHC ring, and the y-axis points upward. 
Cylindrical coordinates $(r,\phi)$ are used in the transverse $(x,y)$ plane, $\phi$ being the 
azimuthal angle around the beam direction. The pseudorapidity is defined in terms of the
polar angle $\theta$ as $\eta = -\textrm{ln}[ \textrm{tan}(\theta/2)]$. 
} and nearly $4\pi$ coverage in solid angle.
A set of tracking detectors around the collision point (collectively referred to as the inner detector) is
located within a superconducting solenoid providing a $2$~T axial magnetic field, and is surrounded by a calorimeter system and a muon spectrometer.
The inner detector (ID) consists of a silicon pixel detector and a 
silicon microstrip tracker, together  providing precision tracking in the pseudorapidity range 
$|\eta| < 2.5$, complemented by a straw-tube transition radiation tracker providing
tracking and electron identification information for $|\eta| < 2.0$.
The electromagnetic calorimeter covers the region $|\eta|<3.2$ and is based on a lead/liquid-argon (LAr) sampling technology. 
The hadronic calorimeter uses a steel/scintillator-tile sampling detector in the region $|\eta|<1.7$ and a copper/LAr detector in the region $1.5 < |\eta| < 3.2$. 
The most forward region of ATLAS, $3.1 <|\eta| < 4.9$, is equipped with 
a forward calorimeter, measuring electromagnetic and hadronic energies in copper/LAr and tungsten/LAr modules.
The muon spectrometer (MS) comprises separate trigger 
and high-precision tracking chambers to measure the deflection of muons in a magnetic field generated by three 
large superconducting toroids with coils arranged with an eightfold azimuthal symmetry around the calorimeters. 
The high-precision chambers cover the range of $|\eta|< 2.7$ with three layers of monitored drift tubes, complemented 
by cathode strip chambers in the forward region, where the particle flux is highest. 
The muon trigger system covers the range $|\eta|< 2.4$ with 
resistive-plate chambers in the barrel and thin-gap chambers in the endcap regions.
A two-level trigger system~\cite{Aaboud:2016leb} is used to select events in real time. It consists of a hardware-based 
first-level trigger that uses a subset of detector information to reduce the event rate to approximately 100 kHz, 
and a software-based high-level trigger system that reduces the event rate to about 1 kHz.
The latter employs algorithms similar to those used offline to identify electrons, muons, photons and jets.

%-------------------------------------------------------------------------------
\section{Phase space for cross-section measurement}
\label{sec:phase_space}
%-------------------------------------------------------------------------------

The fiducial \wz\ cross section is measured in a phase space chosen to follow closely the event selection criteria described in Section~\ref{sec:selection}.
It is based on the kinematics of particle-level objects as defined in Ref.~\cite{TruthWS}. 
These are final-state prompt\footnote{A prompt lepton is a lepton that is not produced in the decay of a hadron, a $\tau$-lepton, or their descendants.} leptons
associated with the $W$ and $Z$ boson decays. 
Charged leptons after QED final-state radiation are ``dressed'' 
by adding to the lepton four-momentum the contributions from photons with an angular distance $\Delta R \equiv \sqrt{(\Delta\eta)^2 + (\Delta\phi)^2} < 0.1$ 
from the lepton. Dressed leptons, and final-state neutrinos that do not originate from hadron or $\tau$-lepton decays, 
are matched to the $W$ and $Z$ boson decay products using 
an algorithm that does not depend on details of the Monte Carlo (MC) generator,
called the ``resonant shape'' algorithm. 
This algorithm is  based on the value of an estimator expressing
the product of the nominal line-shapes of the $W$ and $Z$ resonances
\begin{equation*}\label{eq:Pk}
   P = \left| \frac{1}{ m^2_{(\ell^+,\ell^-)} - \left(m_Z^{\textrm{PDG}}\right)^2 + \textrm{i} \; \Gamma_Z^{\textrm{PDG}} \; m_Z^{\textrm{PDG}} } \right|^2  \times \; \left| \frac {1} { m^2_{(\ell',\nu_{\ell'})} - \left(m_W^{\textrm{PDG}}\right)^2 + \textrm{i} \; \Gamma_W^{\textrm{PDG}} \; m_W^{\textrm{PDG}} } \right|^2 \, ,
 \end{equation*}
where $m_Z^{\textrm{PDG}}$ ($m_W^{\textrm{PDG}}$) and $\Gamma_Z^{\textrm{PDG}}$ ($\Gamma_W^{\textrm{PDG}}$) 
are the world average mass and total width of the $Z$ ($W$) boson, respectively, as reported by the Particle Data Group~\cite{PhysRevD.98.030001}.
The input to the estimator is the invariant mass $m$ of all possible pairs ($\ell^+,\ell^-$) and ($\ell',\nu_{\ell'}$)
satisfying the fiducial selection requirements defined in the next paragraph.
The final choice of which leptons are assigned to the $W$ or $Z$ bosons corresponds 
to the configuration  exhibiting the largest value of the estimator.
Using this specific association algorithm, the gauge boson kinematics can be computed using the 
kinematics of the associated leptons independently of any internal MC generator details.

The reported cross sections are measured in a fiducial phase space
defined at particle level as follows.
The dressed leptons from $Z$ and $W$ boson decay must have $|\eta| < 2.5$ and transverse momentum \pt\ above $15$~\GeV\ and  $20$~\GeV, respectively;
the invariant mass of the two leptons from the $Z$ boson decay differs by at most $10$~\GeV{} 
from the world average value of the $Z$ boson mass $m_Z^{\textrm{PDG}}$. 
The $W$ transverse mass, defined as $m_{\textrm{T}}^{W} = \sqrt{2 \cdot p_{\mathrm{T}}^\nu \cdot p_{\mathrm{T}}^\ell \cdot [1 -\cos{\Delta \phi(\ell, \nu)}]}$, where $\Delta \phi(\ell, \nu)$ 
is the angle between the lepton and the neutrino in the transverse plane, and $\pt^\ell$ and $\pt^\nu$ are the transverse momenta of the lepton from $W$ boson decay and of the neutrino, respectively, must be greater than $30$~\GeV. 
In addition, it is required that the angular distance $\Delta R$ 
between the charged leptons from the $W$ and $Z$ decay is larger than $0.3$, and that
$\Delta R$ between the two leptons from the $Z$ decay is larger than $0.2$.
A requirement that the transverse momentum of the leading lepton be above $27$~\GeV\ reduces the acceptance of the fiducial phase space by less than $0.5\%$. 
This criterion is therefore not added to the definition of the fiducial phase space, while it is present in the selection at the detector level.

The fiducial cross section is extrapolated to the total phase space and corrected for the leptonic 
branching fractions of the $W$ and $Z$ bosons, $(10.86 \pm 0.09) \%$ and $(3.3658 \pm 0.0023) \%$~\cite{PhysRevD.98.030001}, respectively. The total phase space is defined by requiring the 
invariant mass of the lepton pair associated with the $Z$ boson to be in the range $66 < m_{\ell\ell} < 116$~\GeV.

For the differential measurements related to jets, particle-level jets are reconstructed from stable particles 
with a lifetime of $\tau>30$~ps in the simulation.
Stable particles are taken after parton showering, hadronisation, and the decay of particles with $\tau<30$~ps.
Muons, electrons, neutrinos and photons associated with $W$ and $Z$ decays are excluded from the jet collection. The particle-level jets are reconstructed with the anti-$k_{t}$ 
algorithm~\cite{antikt} with a radius parameter $R=0.4$ and are required to have a \pt above $25$~\GeV{} and an absolute value of pseudorapidity below $4.5$.

%-------------------------------------------------------------------------------
\section{Signal and background simulation}
\label{sec:samples}
%-------------------------------------------------------------------------------

A sample of simulated \wz\ events is used to correct the signal yield for detector effects, to extrapolate 
from the fiducial to the total phase space, and to compare the measurements with the theoretical predictions.
The production of \wz\ pairs and the subsequent leptonic decays of the vector bosons were simulated at NLO in QCD using the 
\POWHEG-\textsc{Box}~v2~\cite{powheg1:2004, powheg, Alioli:2010xd, Melia:2011tj} generator, 
interfaced to \PYTHIA~8.210~\cite{Sjostrand:2014zea} for simulation of parton showering, hadronisation and the underlying event. 
Final-state radiation resulting from QED interactions is simulated using  \PYTHIA~8.210 and the AZNLO~\cite{AZNLO:2014} 
set of tuned parameters. 
The CT10~\cite{CT10} PDF set was used for the hard-scattering process, 
while the CTEQ6L1~\cite{Pumplin:2002vw} PDF set was used for the parton shower. 
The sample was generated with dynamic renormalisation and factorisation QCD scales, $\mu_\textrm{R}$ and $\mu_\textrm{F}$, equal to 
$m_{WZ}/2$, where $m_{WZ}$ is the invariant mass of the $WZ$ system. 
An additional \wz\ sample was generated by interfacing \POWHEG-\textsc{Box}~v2 matrix elements to the 
\herwigpp~2.7.1~\cite{Bahr:2008pv} fragmentation model and is used to estimate the uncertainty due to the fragmentation modelling.
Also for this sample, the CT10 and CTEQ6L1 PDF sets are used for the matrix elements and the parton showers, respectively, while QED 
final-state radiation is internally modelled in \herwig.
An alternative signal sample was generated at NLO QCD using the \Sherpa~2.2.2 generator~\cite{Sherpa}.
Matrix elements contain all diagrams with four electroweak vertices. They were calculated for up to one parton 
at NLO and up to three partons at LO using Comix~\cite{Gleisberg:2008fv} and OpenLoops~\cite{Cascioli:2011va}, 
and merged with the \Sherpa parton shower~\cite{Schumann:2007mg} according to the ME+PS$@$NLO prescription~\cite{Hoeche:2012yf}.
The NNPDF3.0nnlo~\cite{Ball:2014uwa} PDF set was used in conjunction with the dedicated parton shower tuning developed by the \Sherpa authors. 
A calculation using \Sherpa~2.1 with one to three partons at LO is also used for comparisons to measured jet observables.
Finally, the NLO QCD predictions from the \mcnlo~v4.0~\cite{Frixione:2002ik}
MC generator interfaced to the \herwig fragmentation model, using the CT10 PDF set, are also used to estimate signal modelling uncertainties.

NNLO QCD cross sections for \wz production in the fiducial and total phase spaces are provided by the 
\matrix computational framework~\cite{Grazzini:2016swo,Grazzini:2017ckn,Grazzini:2017mhc,Cascioli:2011va,Denner:2016kdg,Gehrmann:2015ora,Catani:2012qa,Catani:2007vq}.
The calculation includes contributions from off-shell electroweak bosons and all relevant interference effects.
The renormalisation and factorisation scales were fixed to $(m_Z+m_W)/2$, chosen following Ref.~\cite{Grazzini:2016swo}, 
and the NNPDF3.0nnlo PDF set was used.
The predictions from the \powhegpythia sample were rescaled by a global factor of $1.18$ to match the \NNLO cross section predicted by \MATRIX. 

The background sources in this analysis include processes with two or more electroweak gauge bosons, namely $ZZ$, 
$WW$ and $VVV$ ($V=W,Z$); processes with top quarks, such as $t\bar{t}$ and $t\bar{t}V$, single top and $tZ$; and processes with 
gauge bosons associated with jets or photons ($Z+j$ and $Z\gamma$).
MC simulation is used to estimate the contribution from 
background processes with three or more prompt leptons. Background processes with at least one misidentified 
lepton are evaluated using data-driven techniques and simulated events are used to assess the systematic uncertainties of these backgrounds (see Section~\ref{sec:background}).

The \Sherpa~2.2.2 event generator was used to simulate $q\bar{q}$-initiated $ZZ$ processes
with up to one parton at NLO and up to three partons at LO and using the NNPDF3.0nnlo PDF set.
A \Sherpa~2.1.1 $ZZ$ sample was generated with the loop-induced $gg$-initiated process simulated at LO using the CT10 PDF, 
with up to one additional parton. A $K$-factor of $1.67\pm0.25$ was applied to the cross section of the loop-induced $gg$-initiated
process to account for the NLO corrections~\cite{Caola:2015psa}.
Triboson events were simulated at LO with the \Sherpa~2.1.1 generator using the CT10 PDF set.
The $t\bar{t}V$ processes were generated at NLO with the \textsc{MadGraph5\_aMC@NLO}~\cite{Alwall:2014hca} MC generator
using the NNPDF3.0nlo PDF set interfaced to the \PYTHIA~8.186~\cite{Sjostrand:2007gs} parton shower model.
Finally, the  $tZ$ events were generated at LO with the \textsc{MadGraph5\_aMC@NLO} using the 
NNPDF2.3lo~\cite{Ball:2012cx} PDF set interfaced with \PYTHIA~6.428~\cite{Sjostrand:2006za}. 

All generated MC events were passed through the ATLAS detector simulation~\cite{Aad:2010ah}, based on 
GEANT4~\cite{Agostinelli:2002hh}, and processed using the same reconstruction software as used for the data.
The event samples include the simulation of additional proton--proton interactions (pile-up) 
generated with \PYTHIA~8.186 using the MSTW2008LO~\cite{Martin:2009iq} PDF set 
and the A2~\cite{ATL-PHYS-PUB-2012-003} set of tuned parameters for the underlying event and parton fragmentation. 
Simulated events were reweighted to match the pile-up conditions observed in the data.
Scale factors are applied to simulated events to correct for small differences in the efficiencies observed in data 
and predicted by MC simulation for the trigger, reconstruction, identification, isolation and impact parameter 
requirements of electrons and muons~\cite{ATLAS-CONF-2016-024,PERF-2015-10}. 
Furthermore, the electron energy and muon momentum in simulated events are smeared to account for small 
differences in resolution between data and MC events~\cite{Aaboud:2018ugz,PERF-2015-10}.

%-------------------------------------------------------------------------------
\section{Event selection}
\label{sec:selection}
%-------------------------------------------------------------------------------

Only data recorded with stable beam conditions and with all relevant detector subsystems operational are considered.
Candidate events are selected using triggers~\cite{Aaboud:2016leb} that require at least one electron or muon. 
The transverse momentum threshold applied to leptons by the triggers
in $2015$ was $24$~\GeV\ for electrons and $20$~\GeV\ for muons satisfying a loose isolation requirement based only on ID track information.
Due to the higher instantaneous luminosity in $2016$ the trigger threshold was increased to $26$~\GeV\ for
both the electrons and muons. Furthermore, tighter lepton isolation and tighter electron identification requirements were applied in $2016$.
Possible inefficiencies for leptons with large transverse momentum are reduced by using additional triggers with tighter thresholds, $\pT = 60$~\GeV{} and $50$~\GeV{} for electrons and muons respectively, and no isolation requirements. 
Finally, a single-electron trigger
requiring $\pT>120$~\GeV{} (in $2015$) and  $\pT>140$~\GeV{} (in $2016$) with less restrictive electron
identification criteria was used to increase the selection efficiency for high-$\pt$ electrons.
The combined efficiency of these triggers is close to $100\%$ for \wz events passing the offline selection criteria.

Events are required to have a primary vertex compatible with the luminous region of the LHC.
The primary vertex is defined as the reconstructed vertex with at least two charged particle tracks, that has the largest sum of the $\pT^{2}$ for the associated tracks. 

All final states with three charged leptons (electrons $e$ or muons $\mu$) and missing transverse momentum (\met) from \wz\ leptonic decays are considered. In the following, the different final states are referred to as $\mu^{\pm}\mu^{+}\mu^{-}$, $e^{\pm}\mu^{+}\mu^{-}$, $\mu^{\pm}e^{+}e^{-}$ and $e^{\pm}e^{+}e^{-}$, where the first label is from the charged lepton of the $W$ decay, and the last two labels are for the $Z$ decay.
No requirement on the number of jets is applied.

Muon candidates are identified by tracks reconstructed in the muon spectrometer (MS) and matched to tracks 
reconstructed in the inner detector (ID). Muons are required to pass a ``medium'' identification selection, 
which is based on requirements on the number of hits in the ID and the MS~\cite{PERF-2015-10}. 
The efficiency of this selection averaged over \pt and $\eta$ is larger than $98\%$. 
The muon momentum is calculated by combining the MS measurement, corrected for the energy deposited in 
the calorimeters, and the ID measurement.
The $\pt$ of the muon must be greater than $15$~\GeV{} and its pseudorapidity must satisfy $|\eta|<2.5$.

Electron candidates are reconstructed from energy clusters in the electromagnetic calorimeter matched 
to ID tracks. Electrons are identified using a discriminant that is the value of a likelihood 
function constructed with information about the shape of the electromagnetic showers in the calorimeter, 
the track properties, and the quality of the  track-to-cluster matching for the candidate~\cite{ATLAS-CONF-2016-024}. 
Electrons must satisfy a ``medium'' likelihood requirement, which provides an overall identification efficiency of $90\%$.
The electron momentum is computed from the cluster energy and the direction of the track. The $\pt$ of 
the electron must be greater than $15$~\GeV{} and the pseudorapidity of the cluster must satisfy 
$\abseta< 1.37$ or $1.52 < \abseta < 2.47$ to be within the tracking system, excluding the transition 
region between the barrel and endcap sections of the electromagnetic calorimeter.

Electron and muon candidates are required to originate from the primary vertex. Thus, the significance of 
the track's transverse impact parameter calculated relative to the beam line, $|{d_{0}/\sigma_{d_{0}}}|$, 
must be smaller than $3.0$ for muons and less than $5.0$ for electrons. Furthermore, the longitudinal impact parameter, 
$z_{0}$ (the difference between the value of $z$ of the point on the track at which $d_{0}$ is defined and the 
longitudinal position of the primary vertex), is required to satisfy $|z_0\cdot \sin(\theta)|<0.5$~mm.

Electrons and muons are required to be isolated from other particles using both calorimeter-cluster and ID-track information.
The isolation requirement for electrons is tuned for an efficiency of  at least $90\%$ for $\pT>25$~\GeV\ and 
at least $99\%$ for $\pT>60$~\GeV~\cite{ATLAS-CONF-2016-024}, while fixed requirements on the isolation variables 
are used for muons, providing an efficiency above $90\%$ for $\pT>15$~\GeV\ and at least $99\%$ for $\pT>60$~\GeV~\cite{PERF-2015-10}.

Jets are reconstructed from clusters of energy deposition in the calorimeter~\cite{PERF-2014-07} using the 
anti-$k_{t}$ algorithm~\cite{antikt} with a radius parameter $R=0.4$. 
The energy of jets is calibrated using a jet energy correction derived from both simulation
and {\textit{in situ}} methods using data~\cite{Aaboud:2017jcu}.
Jets with $\pt$ below $60$~\GeV\ and with $|\eta|<2.4$ have to pass a requirement on the 
{\textit{jet vertex tagger}}~\cite{Aad:2015ina}, a likelihood discriminant that uses a combination of 
track and vertex information to suppress jets originating from pile-up activity.
All jets must have $\pt>25$~\GeV{} and be reconstructed in the pseudorapidity range $|\eta|<4.5$. 
Jets in the ID acceptance containing a $b$-hadron are identified with a multivariate algorithm~\cite{PERF-2012-04,ATL-PHYS-PUB-2016-012} that uses the impact parameter and reconstructed secondary vertex information of the tracks contained in the jets.
Jets initiated by $b$-quarks are selected by setting the algorithm's output threshold such that a $70\%$ $b$-jet selection efficiency is achieved in simulated $t\bar{t}$ events. 
The corresponding light-flavour ($u$,$d$,$s$-quark and gluon) and $c$-jet misidentification efficiencies are $0.3$\% and $8.2$\%, respectively.
Corrections to the flavour-tagging efficiencies are based on data-driven calibration analyses.

The transverse momentum of the neutrino is estimated from the missing transverse momentum in the event, \met, 
calculated as the negative vector sum of the transverse momentum of all identified hard physics objects 
(electrons, muons, jets), with a contribution from an additional soft term. 
This soft term is calculated from ID tracks matched to the primary vertex and not assigned to any of the hard objects (electrons, muons and jets)~\cite{Aaboud:2018tkc}.

To avoid cases where the detector response to a single physical object is reconstructed as 
two different final-state objects,  e.g. an electron reconstructed as both an electron and a jet, 
several steps are followed to remove such overlaps, as described in Ref.~\cite{Aaboud:2016lpj}.

Events are required to contain exactly three lepton candidates satisfying the selection criteria described above. 
To ensure that the trigger efficiency is well determined, at least one of the candidate leptons is required to have 
$\pt > 25$~\GeV\ for $2015$ and $\pt > 27$~\GeV\ for $2016$ data, as well as being  geometrically matched to a lepton that was selected by the trigger.

To suppress background processes with at least four prompt leptons, events with a fourth lepton candidate 
satisfying looser selection criteria are rejected. For this looser selection, the lepton $\pt$ requirement is 
lowered to $\pt>5$~\GeV{}, electrons are allowed to be reconstructed in the barrel-endcap calorimeter gap ($1.37< \abseta< 1.52$), and ``loose'' identification requirements~\cite{PERF-2015-10,ATLAS-CONF-2016-024} are used for both the electrons and muons.
A less stringent requirement is applied for electron isolation and is based only on ID track information.

Candidate events are required to have at least one pair of leptons with the same flavour and opposite charge, 
with an invariant mass that is consistent with the nominal \Zboson\ boson mass~\cite{PhysRevD.98.030001} to within $10$~\gev. 
This pair is considered to be the $Z$ boson candidate. If more than one pair can be formed, the pair whose invariant mass 
is closest to the nominal $Z$ boson mass is taken as the \Zboson\ boson candidate. 
The remaining third lepton is assigned to the $W$ boson decay. 
The transverse mass of the $W$ candidate, computed using 
\met and the \pt of the associated lepton, is required to be greater than $30$~\GeV.

Backgrounds originating from misidentified leptons are suppressed by requiring the lepton associated with the $W$ boson 
to satisfy more stringent selection criteria. Thus, the transverse momentum of these leptons is required to be greater 
than $20$~\GeV. Furthermore, charged leptons associated with the $W$ boson decay are required to pass the ``tight''  
identification requirements, which results in an efficiency between $90\%$ and $98\%$ for muons and  
an overall efficiency of $85\%$ for electrons. 
Finally, muons associated to the $W$ boson must also pass a tighter isolation requirement, tuned for an efficiency of at least $90\%$~($99\%$) for $\pT>25~(60)$~\GeV.

%-------------------------------------------------------------------------------
\section{Background estimation}
\label{sec:background}
%-------------------------------------------------------------------------------

The background sources are classified into two groups: events where at least one of the candidate leptons is not a prompt lepton (reducible background) 
and events where all candidates are prompt leptons or are produced in the decay of a $\tau$-lepton (irreducible background). Candidates that are not prompt leptons are also called  ``misidentified'' or ``fake'' leptons.

Events in the first group originate from  $Z+j$, $Z\gamma$, $t\bar{t}$, and $WW$ production processes and constitute about $40\%$ of the total backgrounds.
This reducible background is  estimated with a data-driven method based on the inversion of a matrix containing the efficiencies and the misidentification probabilities for prompt and fake leptons~\cite{ATLASWZ_8TeV,Aad:2014pda}. % 
The method exploits  the classification of the leptons as  loose  or tight  candidates
and  the probability that a fake lepton is misidentified as a loose or tight lepton.
Tight leptons are signal leptons as defined in Section~\ref{sec:selection}.
Loose leptons are leptons that do not meet the isolation and identification criteria of signal leptons but satisfy only looser criteria.
The misidentification probabilities for fake leptons are determined from data using dedicated control samples enriched in misidentified leptons from light- or heavy-flavour jets and from photon conversions.
The lepton efficiencies and misidentification probabilities are combined with event rates in data samples of \wz candidate events where at least one and up to three of the leptons are loose.
Then, solving the system of linear equations, the number of events with at least one misidentified lepton, which represents
 the amount of reducible background in the \wz sample, is obtained.
About $2\%$ of this background contribution arises from events with two fake leptons.
The background from events with three fake leptons, e.g., from multijet processes, is negligible. 
The method allows the shape of any kinematic distribution of reducible background events to be estimated.
Another independent method to assess the reducible background was also considered. 
This method estimates the amount of reducible background using MC simulations scaled to data by process-dependent factors determined from the data-to-MC comparison in dedicated control regions.
Good agreement with the matrix method estimate is obtained at the level of $4$\% in the yield and $40$\% in the shape of the distributions of irreducible background events.

The events contributing to the second group of background processes originate from $ZZ$, $t\bar{t} +V $, $VVV$ (where $V$ = $Z$ or $W$) and  $t Z (j)$ events.
The amount of irreducible  background is estimated using MC simulations because processes with a small cross section and signal leptons can be simulated with a better statistical accuracy than an estimate obtained with data-driven methods.
Events from $VH$ production processes with leptonic decays of the bosons can also contribute. 
This contribution was estimated using MC simulations to be of the order of $0.1$\% and was therefore neglected.
The dominant contribution in this second group is from $ZZ$ production, where one of the leptons from the $ZZ$ decay falls outside the detector acceptance. It represents about $70\%$ of the irreducible background.
The MC-based estimates of the $ZZ$ and $t\bar{t} +V $ backgrounds are validated by comparing
the MC expectations with the event yield and several kinematic distributions of a data sample enriched in $ZZ$ and $t\bar{t} +V $ events, respectively.

The $ZZ$ control sample is selected by requiring a $Z$ candidate that  meets all the analysis selection
criteria and is accompanied by two additional leptons, satisfying the lepton criteria
described in Section~\ref{sec:selection}. 
The $ZZ$ MC expectation needs to be rescaled by a factor of $1.12$ in order to match the observed event yield of data in this control region.
This scaling factor relative to \sherpa predictions is in agreement with the $ZZ$ cross-section measurements performed at $\sqrt{s} = 13$~\TeV~\cite{Aaboud:2017rwm}.
The shapes of distributions of the main kinematic variables are found to be well described by the MC predictions.

The $t\bar{t} +V $ control sample is selected by requiring \wz\ events to have at least two reconstructed $b$-jets.
The $t\bar{t} +V $ MC calculation needs to be rescaled by a factor of $1.3$ in order to match the observed event yield in data.
This scaling factor relative to predictions is in line with the $t\bar{t}V$ cross-section measurements performed at $\sqrt{s} = 13$~\TeV~\cite{Aaboud:2016xve}.
Here again, the distributions of the main kinematic variables are found to be well described by the MC predictions.

%-------------------------------------------------------------------------------
\section{Detector-level results}
\label{sec:RecoResults}
Table~\ref{tab:Results:YieldsSummary} summarises the predicted and observed numbers of events together with the
estimated background contributions.
Only statistical uncertainties are quoted. 
Figure~\ref{fig:Results:WZControlPlots} shows the measured distributions of
the transverse momentum and the invariant mass of the $Z$ candidate,
the transverse mass of the $W$ candidate, and for the $WZ$ system the variable $m_{\mathrm{T}}^{WZ}$, defined as follows: 
\begin{equation*}
m_{\mathrm{T}}^{WZ} = \sqrt{ \left( \sum_{\ell = 1}^3 p_{\mathrm{T}}^\ell + E_{\mathrm{T}}^{\mathrm{miss}} \right)^2
                - \left[ \left(\sum_{\ell = 1}^3 p_x^\ell + E_{x}^{\mathrm{miss}} \right)^2 + \left(\sum_{\ell = 1}^3 p_y^\ell + E_{y}^{\mathrm{miss}} \right)^2 \right]} \, .
\end{equation*}

The \powhegpythia MC prediction is used for the \wz\ signal contribution.
Figure~\ref{fig:Results:WZControlPlots}  indicates that the MC predictions provide a fair description of the shapes of the data distributions.

\begin{table}[!htbp]
\caption{Observed and expected numbers of events after the \wz\ inclusive selection described in Section~\ref{sec:selection} in each of the considered channels and for the sum of all channels.  
The expected number of \wz\ events from \powhegpythia and the estimated number of background events from other processes are detailed.
The \powhegpythia MC prediction is scaled by a global factor of $1.18$ to match the \NNLO cross section predicted by \matrix.
The sum of background events containing misidentified leptons is labelled ``Misid. leptons''.
Only statistical uncertainties are reported.
}
\label{tab:Results:YieldsSummary}
\begin{center}
\begin{tabular}{%
%l
%r@{\,}@{$\pm$}@{\,}r
%r@{\,}@{$\pm$}@{\,}r
%r@{\,}@{$\pm$}@{\,}r 
%r@{\,}@{$\pm$}@{\,}r
%r@{\,}@{$\pm$}@{\,}r
l
S@{$\pm$}@{\,}S
S@{$\pm$}@{\,}S
S@{$\pm$}@{\,}S
S@{$\pm$}@{\,}S
S@{$\pm$}@{\,}S
}
\toprule  
{Channel}  &\multicolumn{2}{c}{$eee$} & \multicolumn{2}{c}{$\mu ee$}  & \multicolumn{2}{c}{$e\mu\mu$} & \multicolumn{2}{c}{$\mu\mu\mu$} & \multicolumn{2}{c}{All} \\
\midrule 
Data 		& \multicolumn{2}{c}{$1279$}  	& \multicolumn{2}{c}{$1281$} & \multicolumn{2}{c}{$1671$}	& \multicolumn{2}{c}{$1929$}	& \multicolumn{2}{c}{$6160$}\\ 
\midrule 
Total Expected 	& 1221 & 7  		& 1281 & 6   	& 1653 & 8		& 1830 & 7      		& 5986 & 14     \\ 
\midrule 
$WZ$ 		& 922 & 5 	    		& 1077 & 6   	& 1256 & 6		& 1523 & 7      		& 4778 & 12     \\ 
Misid. leptons 	& 138 & 5   		& 34 & 2     	& 193 & 5		& 71 & 2	     		& 436 & 8       \\ 
$ZZ$ 		& 86 & 1 	    		& 89 & 1     	& 117 & 1		& 135 & 1       		& 426 & 3       \\ 
$t\bar{t}$+V 	& 50.0 & 0.7    		& 54.0 & 0.7     	& 56.1 & 0.7	& 63.8 & 0.8	        & 225 & 1       \\ 
$tZ$ 		& 23.1 & 0.4 	    	& 24.8 & 0.4     	& 28.8 & 0.4	& 33.5 & 0.5	        & 110 & 1       \\ 
$VVV$ 		&  2.5 & 0.1 	    	&  2.8 & 0.1     	&  3.2 & 0.1	&  3.6 & 0.1	        & 12.0 & 0.2    \\ 
\bottomrule 
\end{tabular}

\end{center}

\end{table}

\begin{figure}[!htbp]

\begin{center}
\includegraphics[width=.4\textwidth]{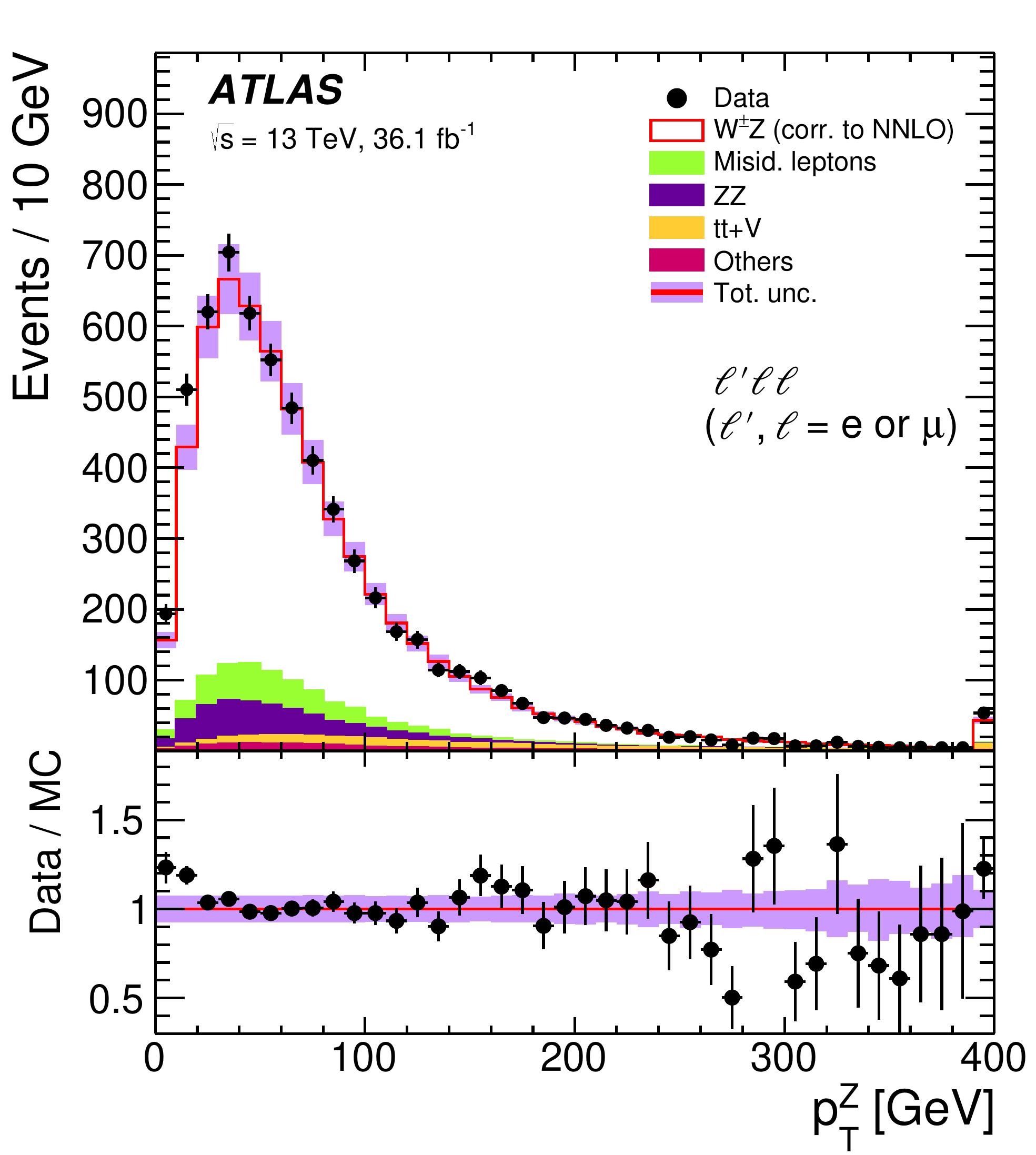}\put(-30,90){{(a)}}
\includegraphics[width=.4\textwidth]{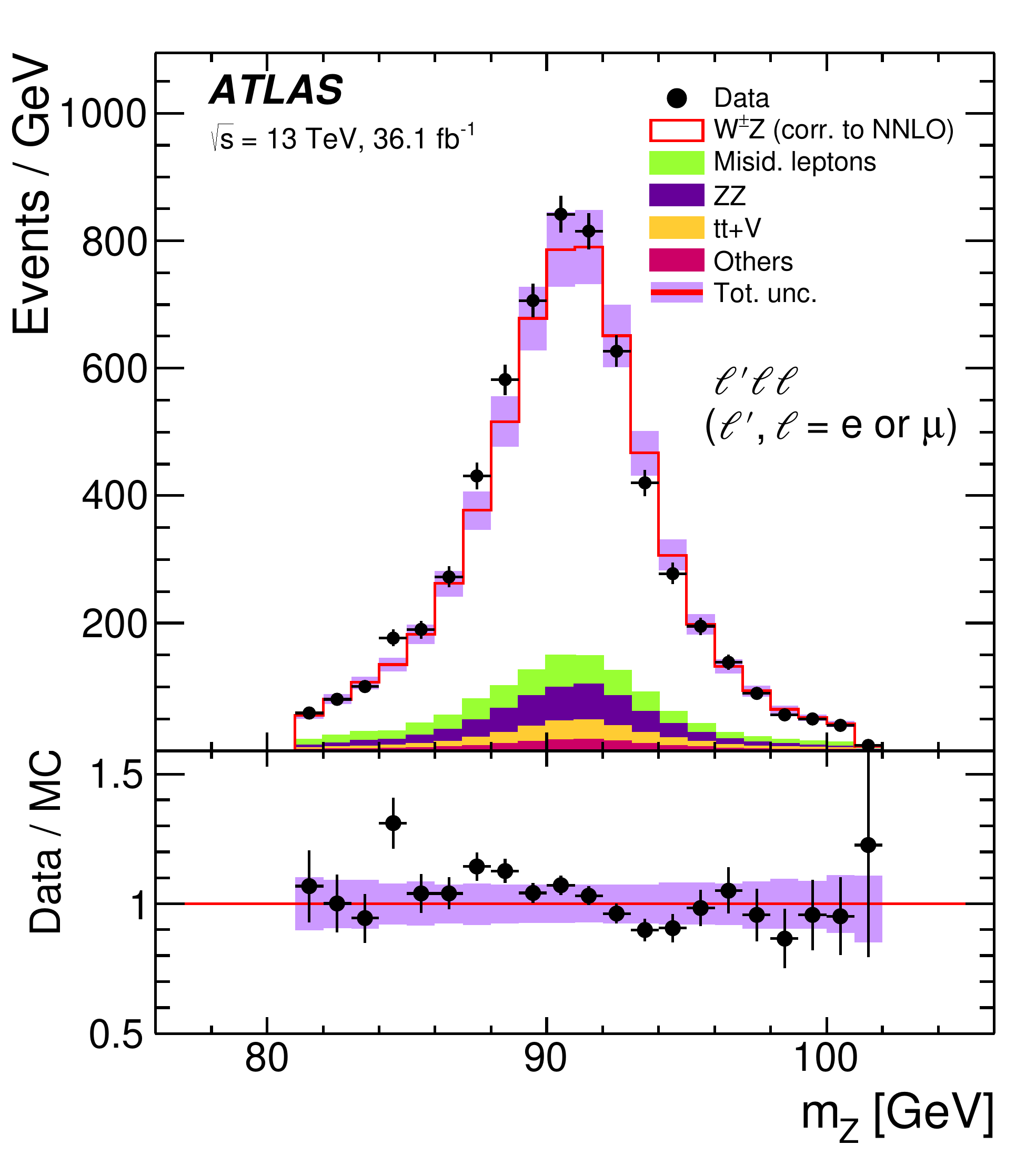}\put(-30,90){{(b)}}\\
\includegraphics[width=.4\textwidth]{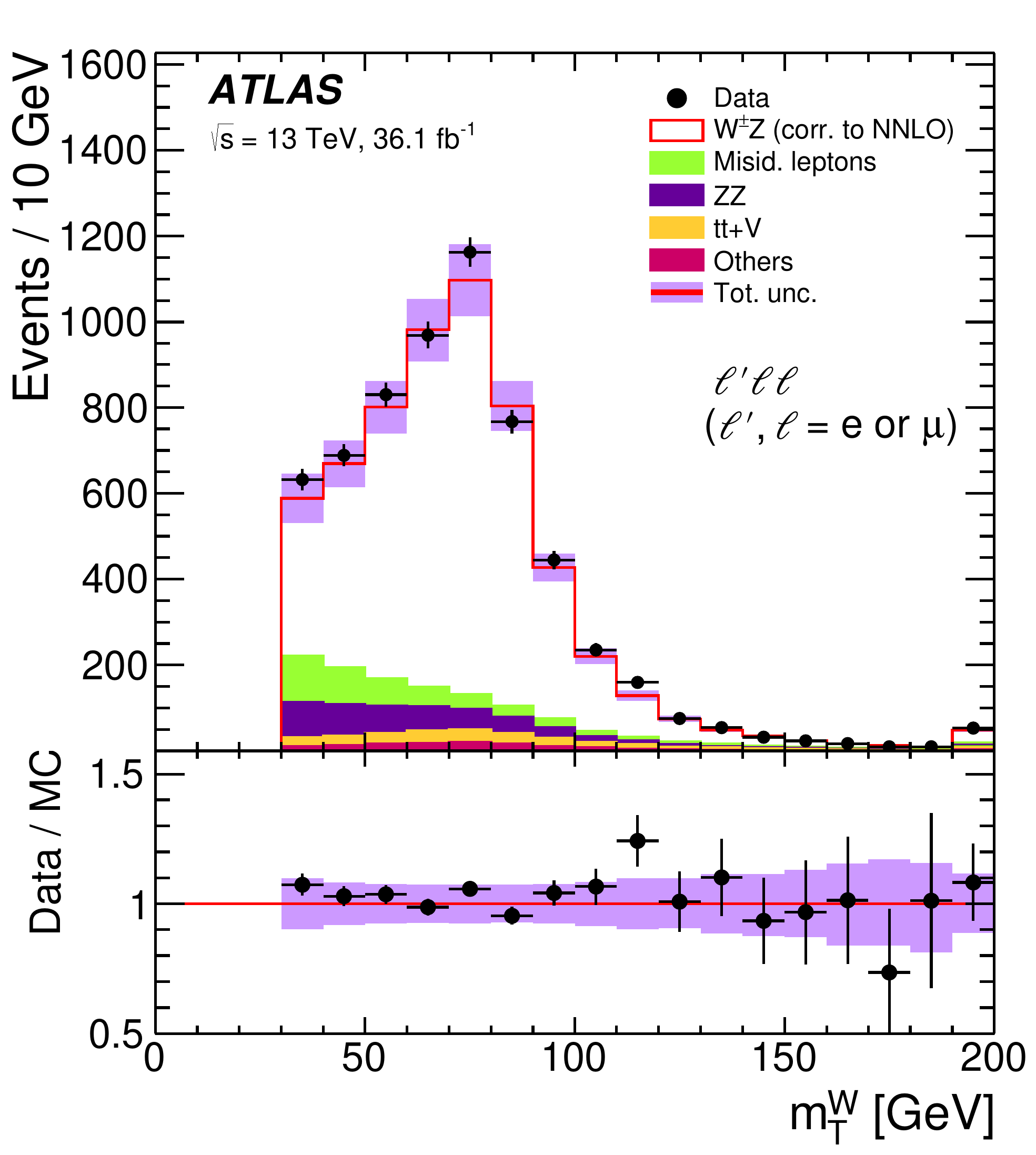}\put(-30,90){{(c)}}
\includegraphics[width=.4\textwidth]{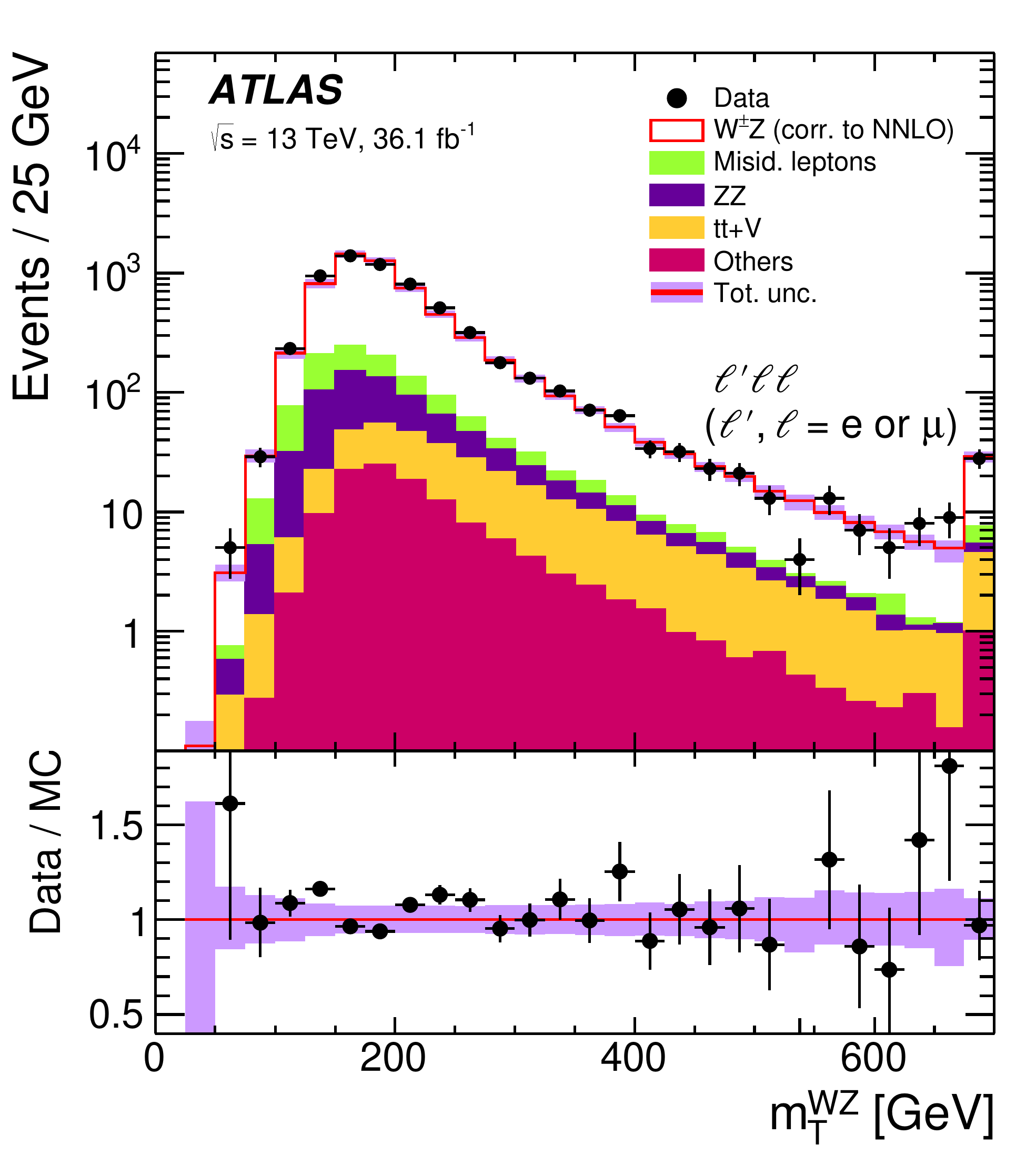}\put(-85,150){{(d)}}\\
\end{center}
 \caption{The distributions, for the sum of all channels, of the kinematic variables (a) $p_\textrm{T}^Z$, (b) $m_Z$, (c) $m_\textrm{T}^W$ and  (d) $m_\textrm{T}^{WZ}$.
The points correspond to the data with the error bars representing the statistical uncertainties, and the histograms correspond to the predictions of the various SM processes. 
The sum of the background processes with misidentified leptons is labelled ``Misid. leptons''.
The \powhegpythia MC prediction is used for the \wz\ signal contribution. 
It is scaled by a global factor of $1.18$ to match the \NNLO cross section predicted by \matrix.
The open red histogram shows the total prediction; the shaded violet band
is the total uncertainty of this prediction.
The last bin contains the overflow.
The lower panels in each figure show the ratio of the data points to the
open red histogram with their respective uncertainties.
}
\label{fig:Results:WZControlPlots}
\end{figure}

%-------------------------------------------------------------------------------
\section{Corrections for detector effects and acceptance}
\label{sec:CrossSectionDefinition}
For a given channel \wz\ $\rightarrow \ell^{'\pm} \nu \ell^+ \ell^-$, where $\ell$ and $\ell^{'}$ indicates each type of lepton ($e$ or $\mu$),  the integrated fiducial cross section that includes the leptonic branching fractions of the
$W$ and $Z$ bosons is calculated as
 \begin{equation*}\label{eq:sigma_fid}
\sigma^{\mathrm{fid.}}_{W^\pm Z \rightarrow \ell^{'} \nu \ell \ell}  = \frac {  N_{\textrm{data}} - N_{\textrm{bkg}} }
 {   \mathcal{L}  \cdot  C_{WZ} }  \times  \left( 1 - \frac {N_\tau} {N_{\textrm{all}} } \right) \, ,
 \end{equation*}
where  $N_{\textrm{data}}$ and  $N_{\textrm{bkg}}$ are the number of observed events and the estimated number of background events, respectively,
$\mathcal{L}$ is the integrated luminosity, and $C_{WZ}$, obtained from simulation, is the ratio of the number of
selected signal events at detector level to the number of events at particle level in the fiducial phase space. This factor corrects for detector efficiencies
and for QED final-state radiation effects. The contribution from $\tau$-lepton decays, amounting approximately to $4\%$, is removed from the
cross-section definition by introducing the term in parentheses.
This term is computed using simulation, where $N_\tau$ is the number of selected
events at detector level in which at least one of the bosons decays into a $\tau$-lepton and $N_{\textrm{all}}$ is the number of selected  $WZ$ events with
decays into any lepton.

The  $C_{WZ}$ factors for $W^- Z$,  $W^+ Z$, and $W^\pm Z$ inclusive processes computed with \powhegpythia for
each of the four leptonic channels are shown in Table~\ref{tab:cwz}.

\begin{table}[!htbp]
\caption{The $C_{WZ}$ factors for each of the $eee$, $\mu{ee}$, ${e}\mu\mu$, and $\mu\mu\mu$ inclusive channels.
The  \powhegpythia MC event sample with the ``resonant shape'' lepton assignment algorithm at particle level
is used. Only statistical uncertainties are reported. }
\label{tab:cwz}
\begin{center}
\begin{tabular}{cccc}
\hline
Channel &   $C_{W^-Z}$ &  $C_{W^{+}Z}$  & $C_{W^\pm Z}$\\
\hline

$eee$     & $0.399$ $\pm$ $0.003$ & $0.394$ $\pm$ $0.003$ & $0.396$ $\pm$ $0.002$ \\
$\mu{ee}$ & $0.470$ $\pm$ $0.004$ & $0.469$ $\pm$ $0.003$ & $0.469$ $\pm$ $0.002$ \\
${e}\mu\mu$  & $0.548$ $\pm$ $0.004$ & $0.541$ $\pm$ $0.003$ & $0.544$ $\pm$ $0.003$ \\
$\mu\mu\mu$ & $0.660$ $\pm$ $0.005$ & $0.663$ $\pm$ $0.004$ &  $0.662$ $\pm$ $0.003$\\
\hline
\end{tabular}
\end{center}

\end{table}

The total cross section is calculated as
\begin{equation*}\label{eq:sigma_tot}
\sigma^{\mathrm{tot.}}_{W^\pm Z}  = \frac{  \sigma^{\mathrm{fid.}}_{W^\pm Z \rightarrow \ell^{'} \nu \ell \ell}  } { \mathcal{B}_W \, \mathcal{B}_Z  \, A_{WZ} } \, ,
 \end{equation*}
where $ \mathcal{B}_W = (10.86 \pm 0.09)\%$ and $\mathcal{B}_Z = (3.3658 \pm 0.0023) \%$ are the $W$ and $Z$ leptonic branching fractions~\cite{PhysRevD.98.030001}, respectively, and
$A_{WZ}$ is the acceptance factor calculated at particle level as the ratio of the number of events in the fiducial
phase space to the number of events in the total phase space as defined in Section~\ref{sec:phase_space}.

A single acceptance factor of $A_{WZ}$ = $0.343$ $\pm$ $0.002$~(stat.), obtained by averaging the acceptance factors computed
in the  $\mu ee$ and $e \mu\mu$ channels, is used
since it was verified for the fiducial phase space used that interference effects related to the presence of
identical leptons in the final state, as in the $eee$ and $\mu \mu \mu$ channels, are below $1\%$ of the cross section.
The use of only the  $\mu{ee}$ and ${e}\mu\mu$ channels for the computation of $A_{WZ}$
avoids the ambiguity arising from the assignment of particle-level final-state leptons to the $W$ and $Z$ bosons.

The differential detector-level distributions within the fiducial phase space are corrected for detector resolution and for QED final-state radiation effects 
using simulated signal events and an iterative Bayesian unfolding method~\cite{DAgostini:1994zf}, as implemented in the RooUnfold toolkit~\cite{RooUnfold}.
The number of iterations used ranges from two to four, depending on the resolution in the unfolded variable.
The width of the bins in each distribution is chosen according to the experimental resolution
and to the statistical significance of the expected number of events in each bin.
The fraction of signal MC events generated in a bin that are reconstructed in the same bin is around $70\%$ on average and always greater than $50\%$, 
except for the jet multiplicity distribution, where it can decrease to $40\%$ for $\njets = 4$.

In the inclusive measurements, the \powhegpythia signal sample is used for unfolding since it provides a fair description of the data distributions.
For differential measurements with jets, the  \SHERPA~2.2.2  signal sample
is used for the computation of the response matrix since this sample includes up to three partons in the matrix element
calculation and therefore better describes the jet multiplicity in data.

%-------------------------------------------------------------------------------
\section{Systematic uncertainties}
\label{sec:systematics}
%-------------------------------------------------------------------------------

The systematic uncertainties in the measured cross sections are due to uncertainties of experimental
and theoretical nature in the acceptance, in the correction procedure for detector effects, in the background estimate and in the luminosity.

The theoretical modelling systematic uncertainties in the $A_{WZ}$ and $C_{WZ}$ factors are 
due to the choice of PDF set, QCD renormalisation 
$\mu_\text{R}$ and factorisation $\mu_\text{F}$ scales, and the parton showering simulation. 
The uncertainties due to the choice of PDF are computed using the CT10 eigenvectors and the envelope
of the differences between the CT10 and CT14~\cite{Dulat:2015mca}, MMHT2014~\cite{HarlandLang:2014zoa} 
and NNPDF 3.0~\cite{Ball:2014uwa} PDF sets, according to the PDF4LHC recommendations~\cite{PDF4LHC}.
The effects of QCD scale uncertainties are estimated by varying $\mu_\text{R}$ and $\mu_\text{F}$ by factors of 
two around the nominal scale $m_{WZ}/2$ with the constraint $0.5 \leq \mu_\text{R} /\mu_\text{F} \leq 2$, 
where $m_{WZ}$ is the invariant mass of the $WZ$ system.
Uncertainties arising from the choice of parton shower model are estimated by interfacing \POWHEG with \PYTHIA or \HERWIG and comparing the results.              
Among these three sources of theoretical uncertainty, only the choice of parton shower model has an effect on the $C_{WZ}$ factors, of $0.5\%$. 
The uncertainty of the acceptance factor $A_{WZ}$ is less than $0.5\%$ due to the PDF choice, less than $0.7\%$ due to the QCD scale choice, and about $0.5\%$ for the choice of parton shower model.

Uncertainties in the unfolding from detector to particle level due to imperfect description of the data by the
MC simulation are evaluated using a data-driven method~\cite{Malaescu:2011yg}. The MC differential distribution is corrected to
match the data distribution and the resulting weighted MC distribution at detector level is unfolded
with the response matrix used in the actual data unfolding. 
The new unfolded distribution is compared with the weighted MC distribution at particle level and the difference is taken as the systematic uncertainty.
Uncertainties in the unfolding are typically of the order of $2$\% but can vary from $0.1$\% to $10$\% depending on the considered observable and bin.
For the inclusive measurements, differences in the unfolded results if the \powhegpythia or \sherpa~2.2.2 MC signal samples are used for the unfolding are found to be covered by these unfolding uncertainties.

The experimental systematic uncertainty on the $C_{WZ}$ factors and on the unfolding 
procedure  includes uncertainties on the scale and resolution of the electron energy, muon momentum, jet energy and \met,
as well as uncertainties on the scale factors applied to the simulation in order to reproduce the
trigger, reconstruction, identification and isolation efficiencies measured in data. 
The systematic uncertainties on the measured cross sections are determined by repeating
the analysis after applying appropriate variations for each source of systematic uncertainty to the simulated samples.
The uncertainties on the jet energy scale and resolution are based on their respective measurements in data~\cite{Aaboud:2017jcu}.
The uncertainty on \MET is estimated by propagating the uncertainties on the transverse momenta of reconstructed objects and
by applying momentum scale and resolution uncertainties to the track-based soft term~\cite{Aaboud:2018tkc}. 
A variation in the pileup reweighting of the MC is included to cover the uncertainty on the ratio between the predicted and measured inelastic cross-section in the fiducial volume defined by $M_X > 13$~\GeV\ where $M_X$ is the mass of the hadronic system~\cite{Aaboud:2016mmw}.
For the measurements of the $W$ charge-dependent cross sections, an uncertainty arising from the charge misidentification of leptons is also considered.
It affects only electrons and leads to an uncertainty of less than $0.15\%$ in the ratio of $W^+Z$ to $W^- Z$ integrated cross sections determined by combining the four decay channels.

The dominant contribution among the experimental systematic uncertainties
in the $eee$ and $\mu ee$  channels is due to the uncertainty on the electron identification efficiency, contributing at most a $2.8\%$ uncertainty to the integrated cross section, while in the $e \mu\mu$ and
$\mu\mu\mu$ channels it originates from the muon reconstruction efficiency   and is at most  $2.8\%$.

The uncertainty on the amount of background from misidentified leptons takes into account the limited number of events in the control regions as well as the differences in background composition between the control region used to determine the lepton misidentification rate and the control regions used to estimate the yield in the signal region.
This results in a total uncertainty of $30\%$ on the misidentified-leptons background yield for the integrated cross-section measurements and of $40\%$ when the shape of the differential distributions of the reducible background events is also considered.

A global uncertainty of $\pm 12\%$ is assigned to the amount of $ZZ$ background predicted by the MC simulation, based on the comparison with data in the $ZZ$ control region. 
Similarly, a global uncertainty of $\pm 30\%$ is assigned to the $t\bar{t} +V $ background.

The uncertainty due to other irreducible background sources is evaluated by propagating the uncertainty in their MC cross sections. These are $20\%$ for $VVV$~\cite{ATL-PHYS-PUB-2016-002} and $15\%$ for $tZ$~\cite{ATLASWZ_8TeV}.

The uncertainty on the combined $2015$+$2016$ integrated luminosity is $2.1\%$. It is derived  from a calibration of the luminosity scale using $x$--$y$ beam-separation scans, following a methodology similar to that detailed in Ref.~\cite{DAPR-2013-01}, and using the LUCID-2 detector for the baseline luminosity measurements~\cite{LUCID2}.
It is applied to the signal normalisation as well as to all background contributions that are estimated 
using only MC simulations and  has an effect of $2.4\%$ on the measured cross sections.

The total systematic uncertainty on the \wz fiducial cross section, excluding the luminosity uncertainty, varies between 
$4\%$ and $6\%$ for the four different measurement channels, and is dominated by the uncertainty on the reducible background estimate. 
Table~\ref{tab:SystematicSummary} shows the statistical uncertainty and the main sources of systematic 
uncertainty on the \wz fiducial cross section for each of the four channels and for their combination.
The modelling uncertainty on the measurements  is dominated by the modelling of the fragmentation.

\begin{table}[!htbp]
\caption{Summary of the relative uncertainties on the measured fiducial cross section $\sigma^{\textrm{fid.}}_{W^\pm Z}$ for each channel and for their combination.
The uncertainties are reported as percentages.
The first rows indicate the main sources of
systematic uncertainty for each channel and their combination, which are treated as correlated between channels.
A row with uncorrelated uncertainties follows, which comprise all uncertainties of statistical origin including MC
statistics as well as statistical uncertainties in the fake-factors calculation, which are uncorrelated between channels.
}
\label{tab:SystematicSummary}
\begin{center}
\begin{tabular}{l r r r r r} 
\hline 
 & $e e e$ & $\mu e e$  & $e \mu\mu$ & $\mu\mu\mu$ & Combined \\ 
\hline 
\multicolumn{6}{c}{Relative uncertainties [\%]}\\ 
\hline 
$e$ energy scale & $0.2$ & $0.1$ & $0.1$ & $<0.1$ & $0.1$ \\ 
$e$ id. efficiency & $2.8$ & $1.8$ & $1.0$ & $<0.1$ & $1.1$ \\ 
$\mu$ momentum scale & $<0.1$ & $<0.1$ & $<0.1$ & $<0.1$ & $<0.1$ \\ 
$\mu$ id. efficiency & $<0.1$ & $1.3$ & $1.6$ & $2.8$ & $1.5$ \\ 
$E_{\mathrm{T}}^{\mathrm{miss}}$ and jets & $0.2$ & $0.2$ & $0.3$ & $0.5$ & $0.3$ \\ 
Trigger & $<0.1$ & $<0.1$ & $0.2$ & $0.3$ & $0.2$ \\ 
Pile-up & $1.0$ & $1.5$ & $1.2$ & $1.5$ & $1.3$ \\ 
Misid. leptons background & $4.7$ & $1.1$ & $4.5$ & $1.6$ & $1.9$ \\ 
$ZZ$ background & $1.0$ & $1.0$ & $1.1$ & $1.0$ & $1.0$ \\ 
Other backgrounds & $1.6$ & $1.5$ & $1.4$ & $1.2$ & $1.4$ \\ 
\hline 
Uncorrelated& $0.7$ & $0.6$ & $0.7$ & $0.5$ & $0.3$ \\ 
\hline 
Total systematic uncertainty& $6.0$ & $3.5$ & $5.4$ & $4.1$ & $3.6$ \\ 
Luminosity& $2.2$ & $2.2$ & $2.2$ & $2.2$ & $2.2$ \\ 
Theoretical modelling & $0.5$ & $0.5$ & $0.5$ & $0.5$ & $0.5$ \\ 
Statistics& $3.6$ & $3.3$ & $3.2$ & $2.7$ & $1.6$ \\ 
\hline 
Total& $7.3$ & $5.3$ & $6.6$ & $5.3$ & $4.5$ \\ 
\hline 
\end{tabular} 

\end{center}
\end{table}

%-------------------------------------------------------------------------------
\section{Cross-section measurements}
\label{sec:CrossSections}
\subsection{Integrated cross sections}
\label{sec:IntCrossSections}
The measured fiducial  cross sections for the four channels are combined using a $\chi^2$ minimisation method that accounts for correlations between the sources of
systematic uncertainty affecting each channel~\cite{Glazov:2005rn,Aaron:2009bp,Aaron:2009wt}.
The combination of the \wz cross sections  in the fiducial phase space for the four channels
yields a $\chi^2$ per degree of freedom ($\mathrm{dof}$) of $\chi^2/n_{\mathrm{dof}} = 3.3/3$.
The combinations of the $W^+ Z$ and $W^- Z$ cross sections separately yield $\chi^2/n_{\mathrm{dof}} = 3.7/3$ and $1.5/3$, respectively.

The \wz production cross section in the fiducial phase space resulting from the combination of the
four channels including the $W$ and $Z$ branching ratio in a single leptonic channel with muons or electrons
is

 \begin{equation*}
\sigma_{W^\pm Z \rightarrow \ell^{'} \nu \ell \ell}^{\mathrm{fid.}} =  63.7~\pm~ 1.0 \, \textrm{(stat.)}~\pm~2.3 \, \textrm{(exp.\, syst.)}~\pm~0.3 \, \textrm{(mod.\, syst)}~\pm 1.4 \, \textrm{(lumi.) fb},
\end{equation*}   

where the uncertainties correspond to statistical, experimental systematic, modelling systematic and luminosity uncertainties, respectively.
The corresponding SM NNLO QCD prediction from \MATRIX is $61.5 ^{+1.4}_{-1.3} \, \mathrm{fb}$, where the uncertainty corresponds to the QCD scale uncertainty estimated conventionally by varying the scales $\mu_{\mathrm{R}}$ and $\mu_{\mathrm{F}}$ by factors of two around the nominal value of $(m_W+m_Z)/2$ with the constraint $0.5 \leq \mu_{\mathrm{R}} /\mu_{\mathrm{F}} \leq 2$. 
This prediction is obtained by correcting the result in Ref.~\cite{Grazzini:2017ckn} for Born level leptons to dressed leptons by a factor of $0.96$, which is estimated in the fiducial phase space using \powhegpythia.
Changing the PDF set used from NNPDF3.0nnlo to MMHT2014 or CT14 affects the \matrix prediction by $+2\%$ and $+1\%$, respectively.
The uncertainty due to varying the \alphas coupling constant value used in the PDF determination is $0.6\%$ and $1.0\%$ for $W^+Z$ and $W^-Z$ production, respectively.
The measured \wz production cross sections are compared with the SM NNLO prediction from \MATRIX using three different PDF sets, 
NNDPF3.0nnlo, MMHT2014 and CT14, as well as with NLO predictions from \sherpa~2.2.2 in Figure~\ref{fig:xsectionperchannel}.
All results for $W^\pm Z$, $W^+Z$ and $W^-Z$ final states are reported in Table~\ref{tab:XSection:FiducialCSAll}.
The NNLO SM calculations reproduce the measured cross sections well.
The production of \wz\ in association with two jets produced as a result of electroweak processes is not included in the NNLO SM prediction and amounts to $1.2\%$ of the measured cross section, as estimated using \sherpa 2.2.2.

\begin{table}[!htbp]
\caption{Fiducial integrated cross section in fb, for $W^\pm Z$, $W^+ Z$ and $W^- Z$ production, measured in each of the channels $eee$, $\mu{ee}$, ${e}\mu\mu$, and $\mu\mu\mu$ and for all four channels combined.
The statistical ($\delta_{\mathrm{stat.}}$), experimental systematic ($\delta_{\mathrm{exp.\, syst.}}$), modelling systematic ($\delta_{\mathrm{mod.\, syst.}}$), luminosity ($\delta_{\mathrm{lumi.}}$) and total ($\delta_{\mathrm{tot.}}$) uncertainties are given in percent.
The NNLO SM predictions from \matrix using the NNDPF3.0nnlo set are also reported.
}
\label{tab:XSection:FiducialCSAll}
\begin{center}
\begin{tabular}{lcccccc}

\toprule
Channel   &  $\sigma^{\mathrm{fid.}}$ & $\delta_{\mathrm{stat.}}$ & $\delta_{\mathrm{exp.\, syst.}}$ & $\delta_{\mathrm{mod.\, syst.}}$& $\delta_{\mathrm{lumi.}}$ &$\delta_{\mathrm{tot.}}$\\
          &  [fb] & [\%] &  [\%] & [\%]& [\%] &[\%]\\
\midrule
\multicolumn{6}{c}{~}\\ [-12.0pt]
\multicolumn{6}{c}{$\sigma^{\mathrm{fid.}}_{W^{\pm} Z \rightarrow \ell^{'} \nu \ell \ell}$}\\ [4.0pt]
\midrule
$e^\pm{ee}$                 &  65.8    &       	3.6 &  6.0 &        0.5  &	       2.2 &  7.3\\   
$\mu^\pm{ee}$               &  61.2    &      	3.3 &  3.5 &        0.5  &	       2.2 & 5.3\\ 
$e^\pm\mu\mu $              &  62.4    &       	3.2 &  5.4 &        0.5  &     	       2.2 &  6.6\\ 
$\mu^\pm\mu\mu$             &  65.3    &        2.7 &  4.1 &        0.5  &	       2.2 &  5.3\\ 
\midrule
Combined      		    &  63.7    &        1.6 &  3.6 &        0.5  &	       2.2 &  4.5\\ 
\midrule 
SM prediction  		    &  61.5    &        --- &  --- &       --- &             --- &  $^{2.3} _{2.1}$\\ 
\midrule 
\multicolumn{6}{c}{~}\\ [-12.0pt]
\multicolumn{6}{c}{$\sigma^{\mathrm{fid.}}_{W^{+} Z \rightarrow \ell^{'} \nu \ell \ell}$}\\ [4.0pt]
\midrule
$e^+ee$                       & 40.8    &          4.6 &            5.4 &           0.5  &	       2.2 &   7.4\\   
$\mu^+{ee}$                   & 36.5    &          4.3 &            3.3 &           0.5  &	       2.2 &   5.8\\ 
$e^+\mu\mu$                   & 36.7    &          4.1 &            5.0 &           0.5  &	       2.2 &   6.8\\ 
$\mu^+\mu\mu$                 & 38.2    &          3.5 &            4.0 &           0.5  &     2.2 &   5.7\\ 
\midrule
Combined         	      & 37.9    &          2.0 &             3.4 &          0.5   &	       2.2 &   4.5\\ 
\midrule 
SM prediction  		      & 36.3    &          --- &  --- &           	--- &	        --- &  	 $^{2.2} _{2.0}$\\ 
\midrule 
\multicolumn{6}{c}{~}\\ [-12.0pt]
\multicolumn{6}{c}{$\sigma^{\mathrm{fid.}}_{W^{-} Z \rightarrow \ell^{'} \nu \ell \ell}$}\\ [4.0pt]
\midrule
$e^-ee$                      & 24.9   &            6.1 &            7.1 &           0.5  &     2.2 &	       9.6\\   
$\mu^-{ee}$                  & 24.8   &            5.3 &            4.0 &           0.5  &     2.2 &	       7.0 \\ 
$e^-\mu\mu$                  & 25.7   &            5.1 &            6.2 &           0.5  &     2.2 &	       8.3\\ 
$\mu^-\mu\mu$                & 27.1   &            4.3 &            4.3 &           0.5  &     2.2 &	       6.4 \\  
\midrule
Combined           	     & 25.9   &             8.1 &            4.0 &          0.5  &     2.2 &	       5.2\\ 
\midrule
SM prediction  		     & 25.2    &          --- &  --- &            --- &			 --- &  $^{2.3} _{2.1}$\\ 
\bottomrule 
\end{tabular}

\end{center}
\end{table}

The ratio of the $ W^+Z$ to $W^-Z$ production cross sections is

\begin{equation*}
\frac{\sigma_{W^{+}Z \rightarrow \ell^{'} \nu \ell \ell}^{\textrm{fid.}}}{\sigma_{W^{-}Z \rightarrow \ell^{'} \nu \ell \ell}^{\textrm{fid.}}}  =  1.47\pm 0.05 \,\textrm{(stat.)} \pm 0.02 \,\textrm{(syst.)}.
\end{equation*}

Most of the systematic uncertainties, especially the luminosity uncertainty, almost cancel out in the ratio, so that the measurement is dominated by the statistical uncertainty.
The measured cross-section ratios, for each channel and for their combination, are compared in Figure~\ref{fig:WPMXSection:WpWmRatio} 
with the SM prediction of $1.44^{+0.03}_{-0.06}$, calculated with \MATRIX~\cite{Grazzini:2017ckn} and the NNDPF3.0nnlo PDF set.
The uncertainties correspond to PDF uncertainties estimated at NLO with \powhegpythia using the CT10 eigenvectors and the envelope of the differences between the CT10 and CT14, MMHT2014 and NNPDF 3.0nnlo PDF sets. 
The effects of QCD scale uncertainties on the predicted cross-section ratio are negligible.
The cross-section ratio is also calculated with \matrix using the MMHT2014 and CT14 PDF sets, yielding values of  $1.42$ and $1.44$, respectively, as shown in Figure~\ref{fig:WPMXSection:WpWmRatio}.

\begin{figure}[!htbp]
\begin{center}
\includegraphics[width=9cm]{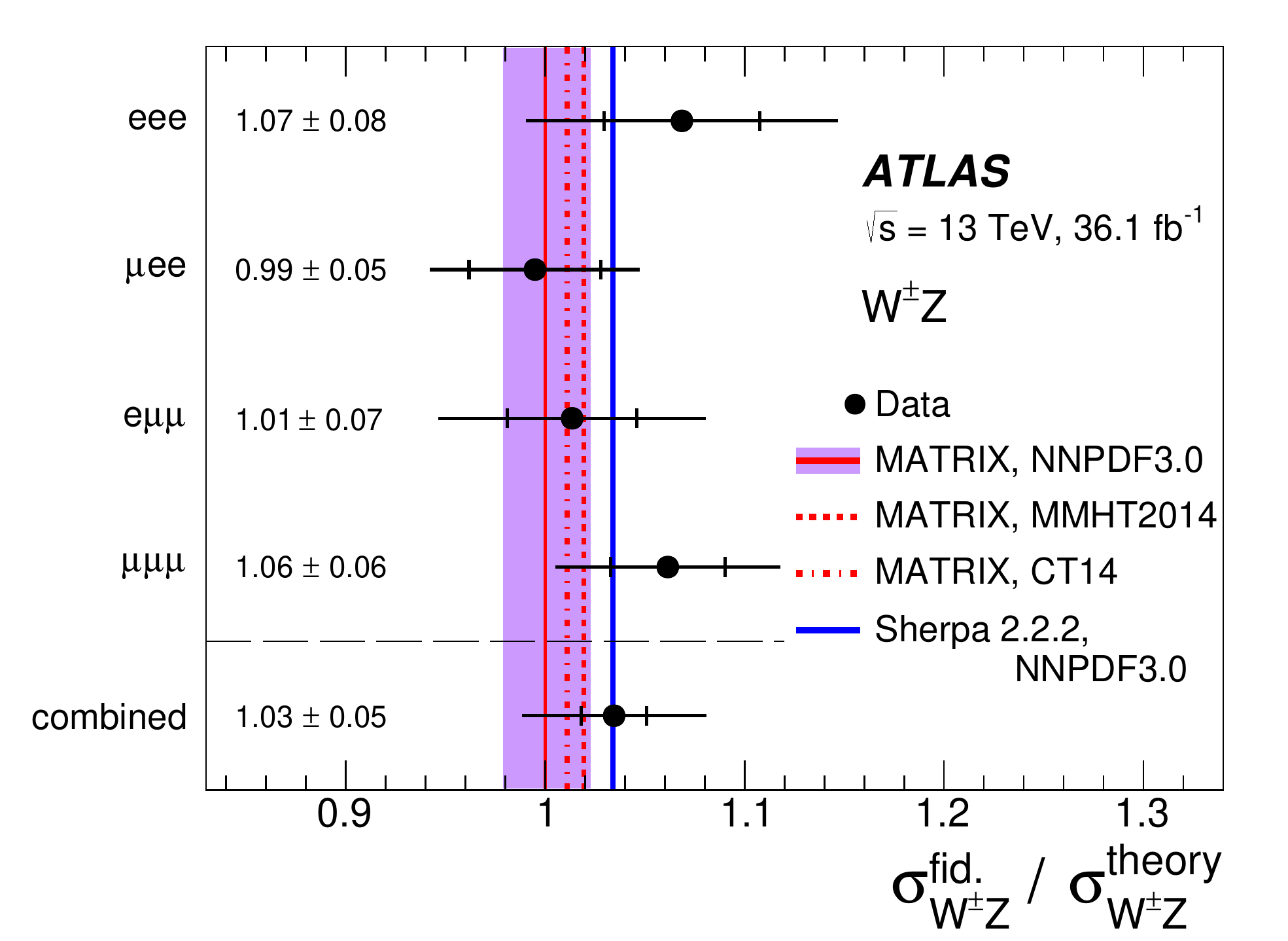}
\caption{Ratio of the measured \wz\ integrated cross sections in the fiducial phase space to the NNLO SM prediction from \MATRIX in each of the four channels and for their combination. 
The inner and outer error bars on the data points represent the statistical and total uncertainties, respectively.
The NNLO SM prediction from \MATRIX using the NNPDF3.0nnlo PDF set is shown as the red line; 
the shaded violet band shows the effect of QCD scale uncertainties on this prediction.
The prediction from \matrix using the MMHT2014 and CT14 PDF sets and the NLO prediction from \sherpa~2.2.2 are also displayed as dashed-red, dotted-dashed-red and blue lines, respectively.
}
\label{fig:xsectionperchannel}
\end{center}
\end{figure}

\begin{figure}[!htbp]
\begin{center}
\includegraphics[width=9cm]{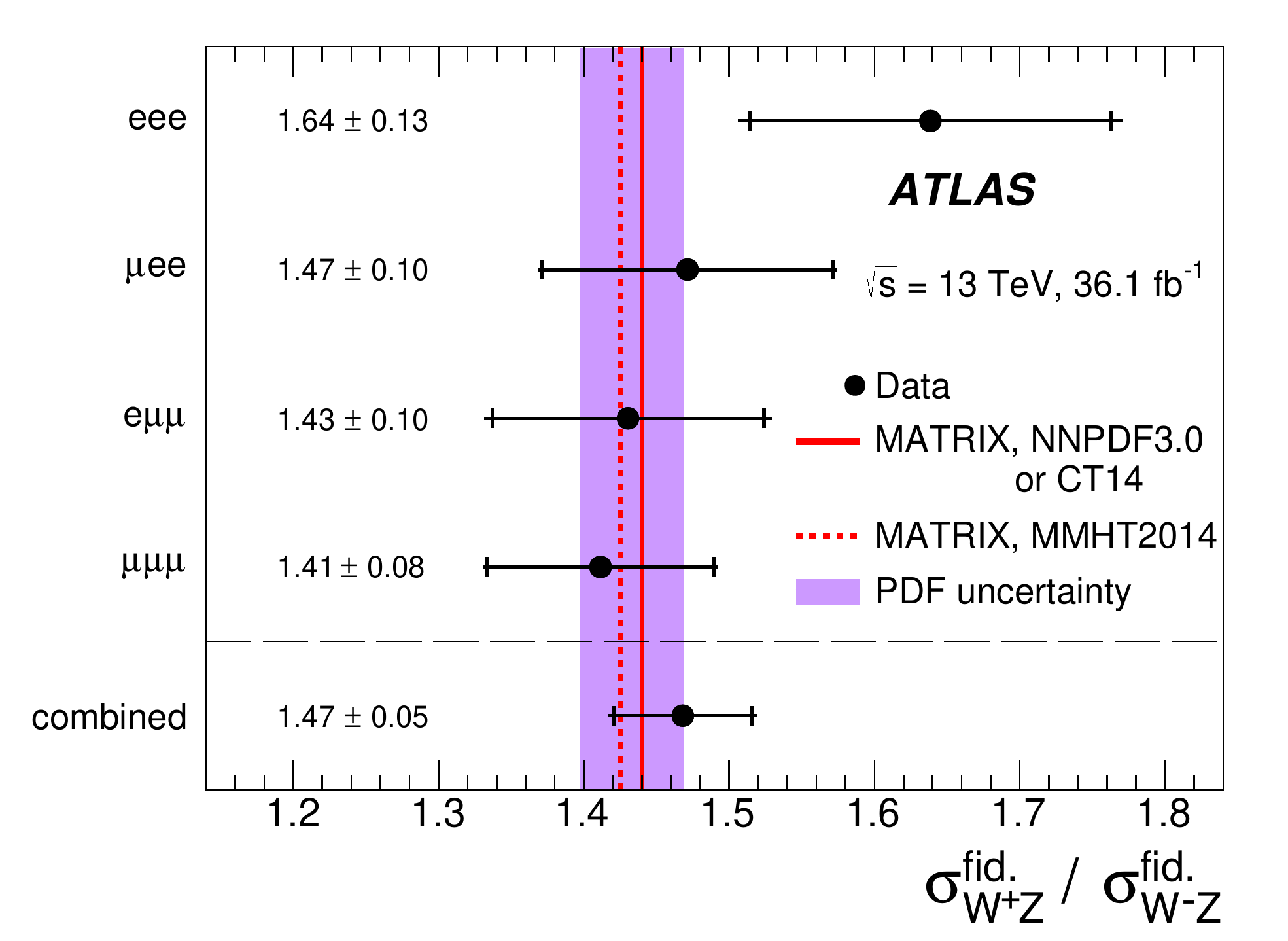}     
\caption{Measured ratio $\sigma^{\mathrm{fid.}}_{W^{+}Z} / \sigma^{\mathrm{fid.}}_{W^{-}Z}$ of $W^{+}Z$ and $W^{-}Z$ integrated cross sections in the fiducial phase space in each of the four channels and for their combination.
The error bars on the data points represent the total uncertainties, which are dominated by the statistical uncertainties.
The \NNLO SM predictions from \MATRIX using the NNPDF3.0nnlo or CT14 PDF sets are equal and represented as a single red line.
The shaded violet band represents the effect of PDF uncertainties estimated using the \powhegpythia NLO calculation using the CT10 eigenvectors and the envelope of the differences between the CT10 and CT14, MMHT2014 and NNPDF 3.0nnlo PDF sets.
The \matrix prediction using the MMHT2014 PDF set is also displayed as the dashed-red line.
}

\label{fig:WPMXSection:WpWmRatio}
\end{center}
\end{figure}

The combined fiducial cross section is extrapolated to the total phase space.
The result is

\begin{equation*}
\sigma_{W^{\pm}Z}^{\mathrm{tot.}}  =  51.0 \pm 0.8 \,\mathrm{(stat.)} \pm 1.8 \,\mathrm{(exp.\, syst.)} \pm 0.9 \,\mathrm{(mod.\, syst.)} \pm 1.1 \,\mathrm{(lumi.) \, pb},
\end{equation*}

where the modelling uncertainty accounts for the uncertainties in the $A_{WZ}$ factor due to the choice of PDF set, QCD scales and the fragmentation model.
The NNLO SM prediction calculated with \matrix~\cite{Grazzini:2016swo} is $49.1 ^{+1.1}_{-1.0} \,\mathrm{(scale)} \; \mathrm{pb}$, which is in  good agreement with the present measurement.
As the \matrix calculation does not include effects of QED final-state radiation, a correction factor of $0.99$, as estimated from \powhegpythia in the total phase space, is applied to it to obtain the above prediction.

\subsection{Differential cross sections}
\label{sec:DiffCrossSections}
For the measurements of the differential distributions, all four decay channels, $eee$, $e\mu\mu$, $\mu ee$,
and $\mu\mu\mu$, are added together.
The resulting distributions are unfolded with a response matrix computed
using a \powhegpythia MC signal sample that includes all four topologies and is divided by four, such
that cross sections refer to final states where the $W$ and $Z$ bosons decay in a single leptonic channel with muons or electrons.

\begin{figure}[!htbp]
\begin{center}
\includegraphics[width=.4\textwidth]{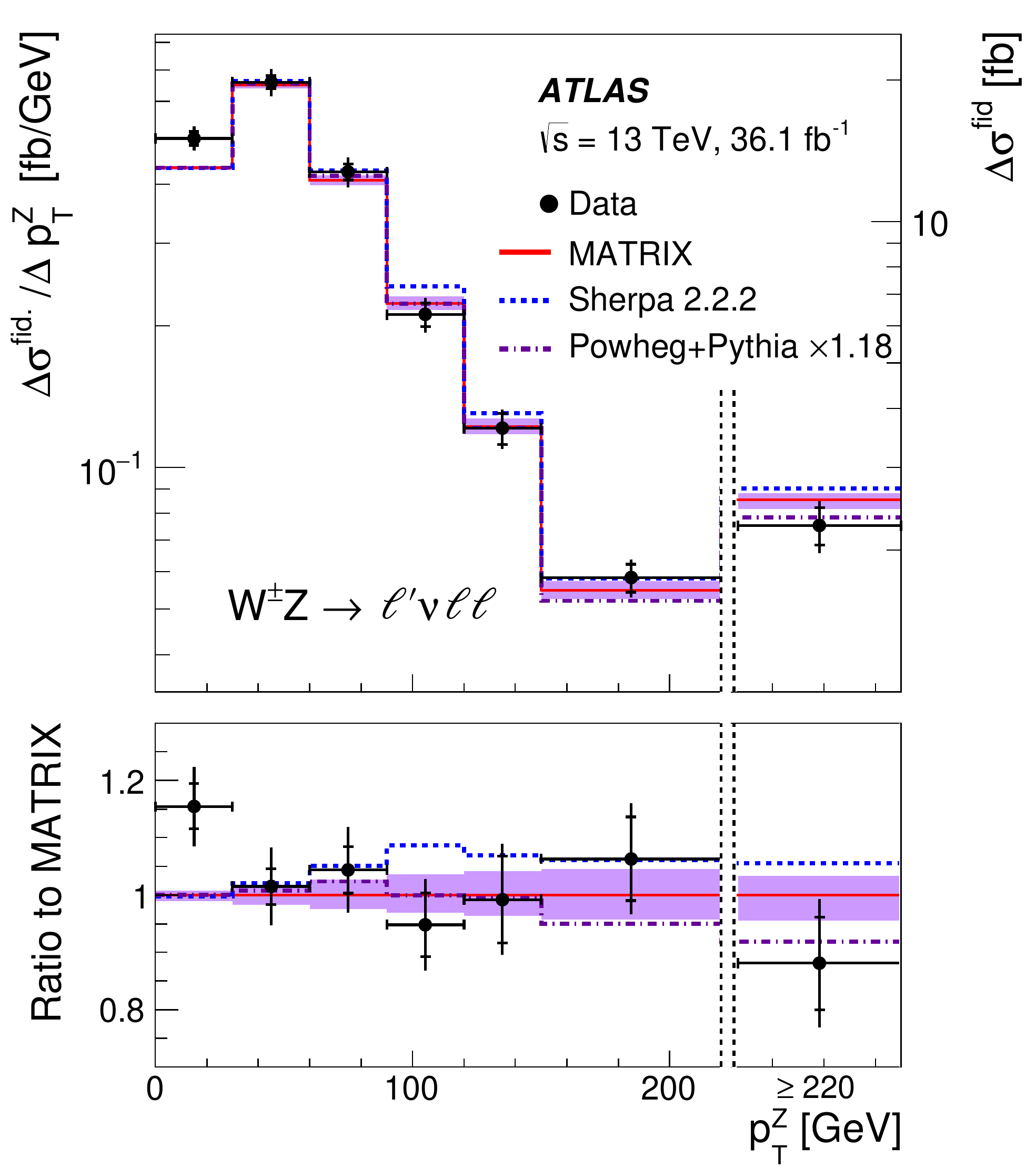}\put(-110,190){{(a)}}\quad
\includegraphics[width=.4\textwidth]{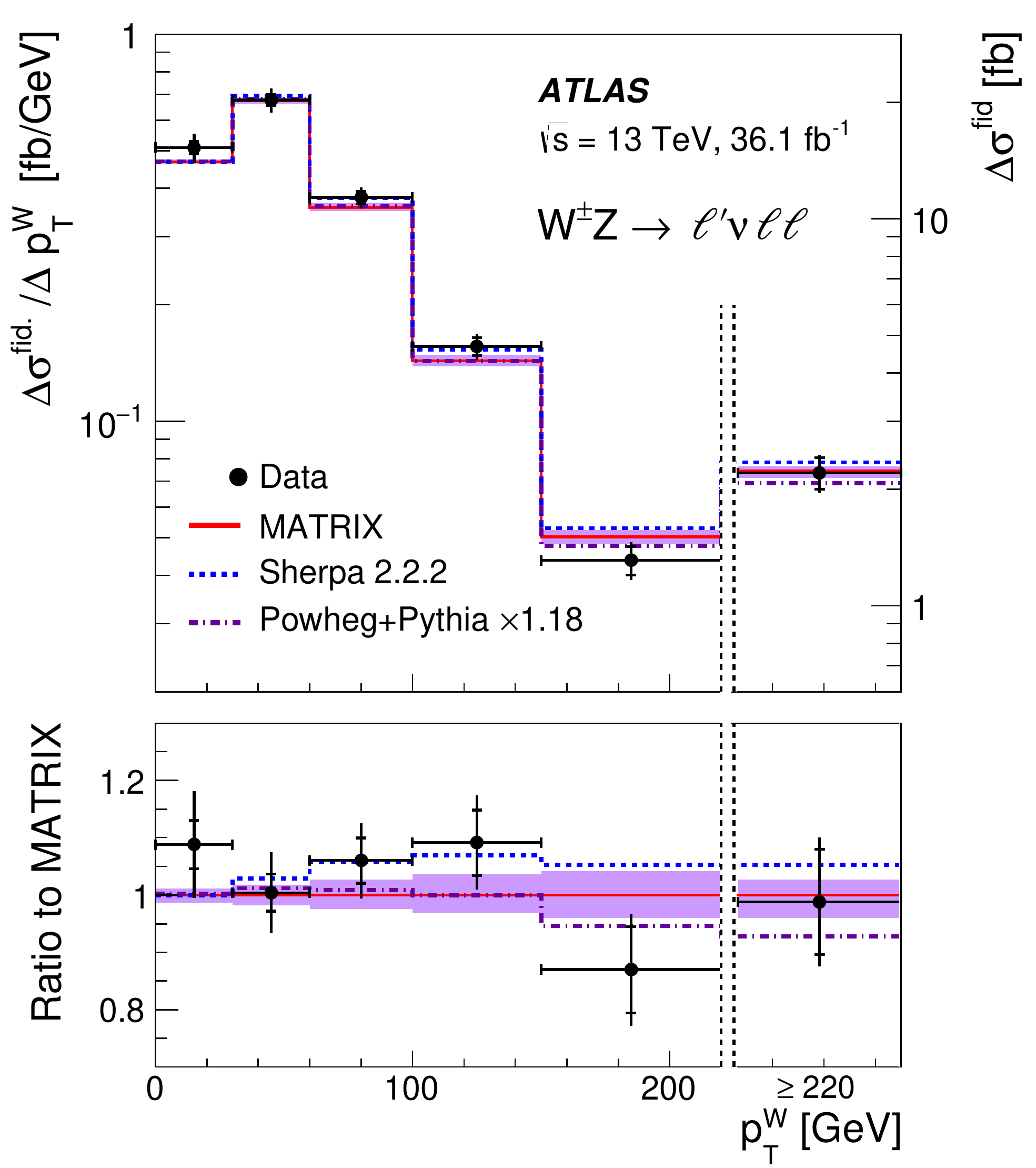}\put(-110,190){{(b)}}\\
\includegraphics[width=0.4\textwidth]{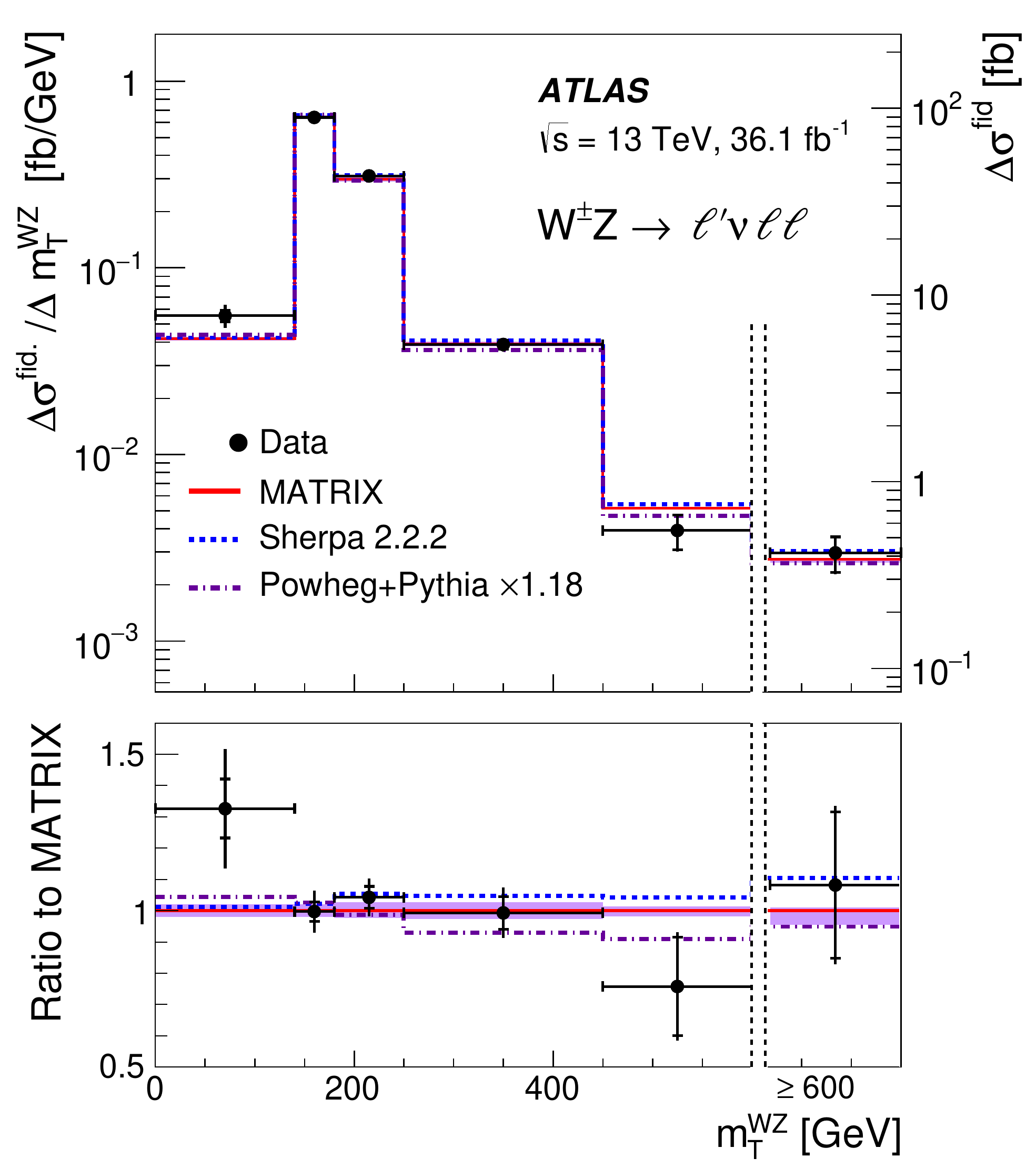}\put(-110,190){{(c)}}\quad
\includegraphics[width=0.4\textwidth]{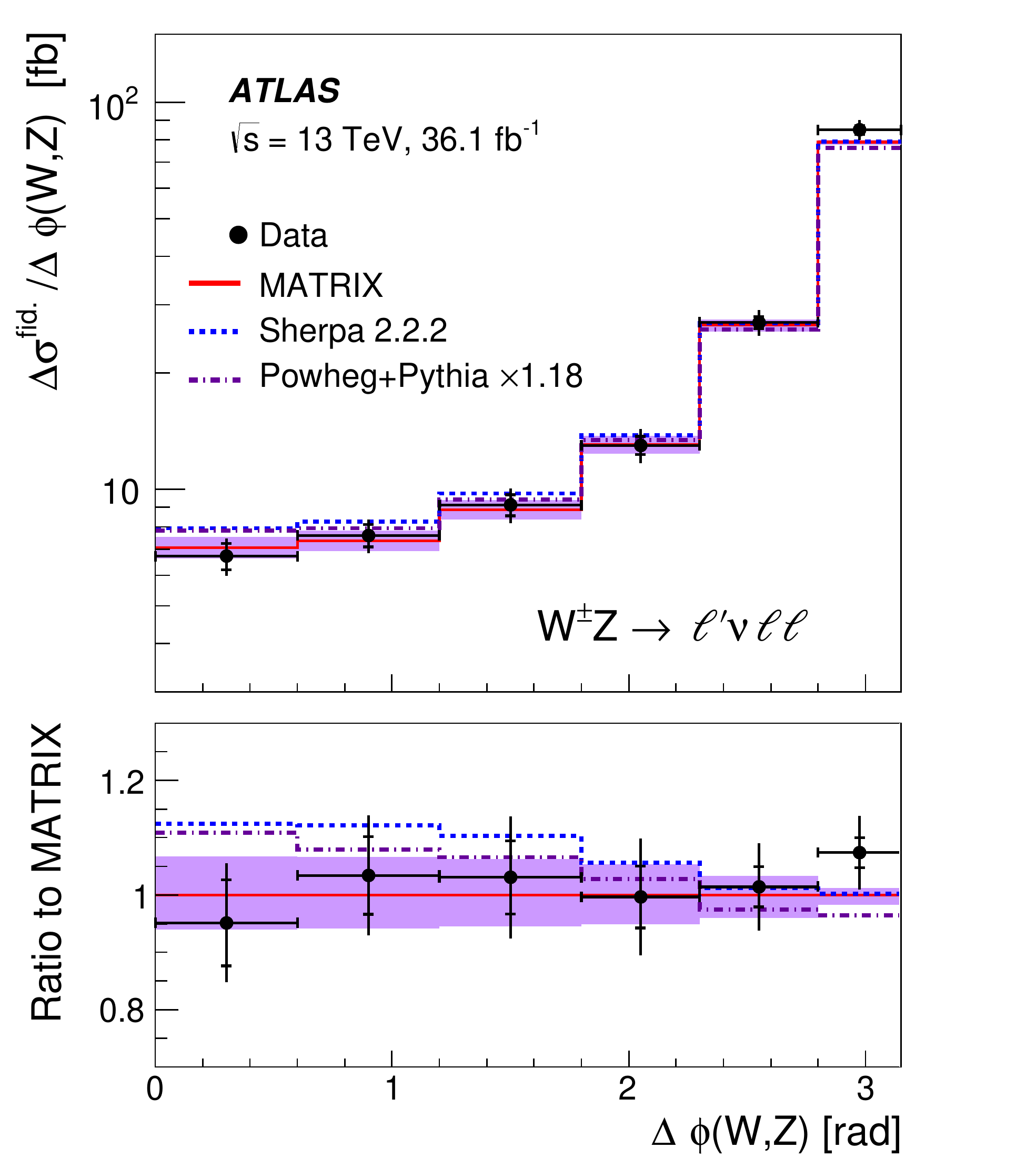}\put(-50,190){{(d)}}
\caption{The measured $W^{\pm}Z$ differential cross section in the fiducial phase space as a function of  (a)    $p_\textrm{T}^Z$,  (b) $p_\textrm{T}^W$, (c) \mtwz  and (d) \dPhiWZ.
The inner and outer error bars on the data points represent the statistical and total uncertainties, respectively.
The measurements are compared with the NNLO prediction from \matrix (red line, see text for details).
The violet band shows how the QCD scale uncertainties affect the NNLO predictions.
The predictions from the \powhegpythia and \sherpa MC generators are also indicated by dotted-dashed and dashed lines, respectively.
In (a), (b) and (c), the right vertical axis refers to the last cross-section point, separated from the others by vertical dashed lines, 
as this last bin is integrated up to the maximum value reached in the phase space and the cross section is not divided by the bin width. }
\label{fig:DiffXSection:pTZW}
\end{center}
\end{figure}

The \wz production cross section is  measured as a function of several variables: the transverse momenta
of the $Z$ and $W$ bosons, $p_\textrm{T}^Z$ and $p_\textrm{T}^W$, 
the transverse mass of the \wz system $m_\textrm{T}^{WZ}$ and the azimuthal angle  between the $W$ and $Z$ bosons in Figure~\ref{fig:DiffXSection:pTZW}; as a function of
the \pt of the neutrino associated with the decay of the $W$ boson, $p_\textrm{T}^\nu$,
and the absolute difference between the rapidities of the $Z$ boson and the charged lepton
from the decay of the $W$ boson, $|y_Z - y_{\ell,W}|$ in Figure~\ref{fig:DiffXSection:yZlW_all}.

In order to derive $p_\textrm{T}^W$ and $p_\textrm{T}^\nu$ from data events, it is assumed that the whole $E_\textrm{T}^\textrm{miss}$ of each event arises from the neutrino of the $W$ boson decay.
The validity of this assumption was verified for SM $WZ$ events using MC samples at the level of precision of the present results.

\begin{figure}[!htbp]
\begin{center}
\includegraphics[width=0.4\textwidth]{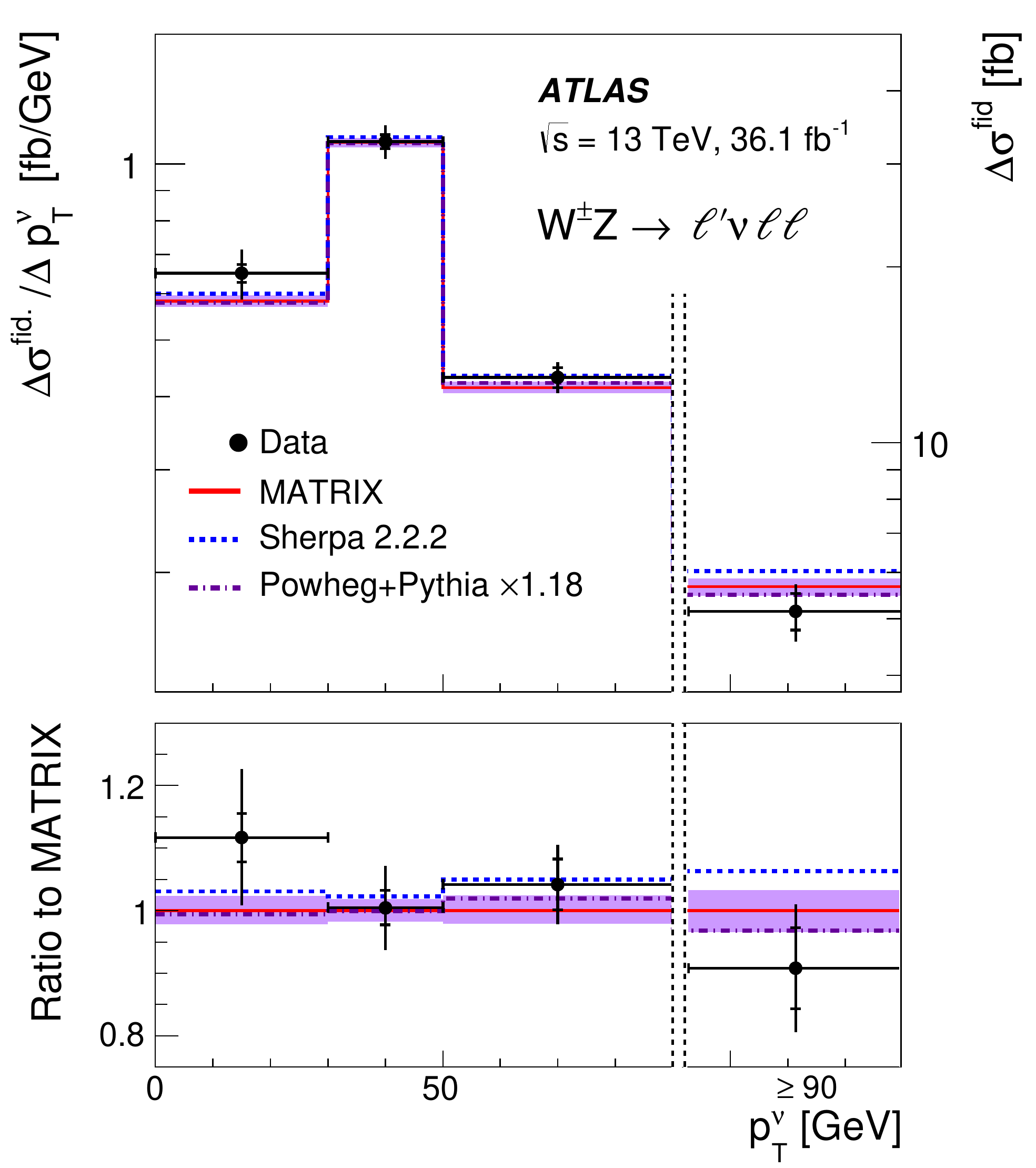}\put(-110,190){{(a)}}\quad
\includegraphics[width=0.4\textwidth]{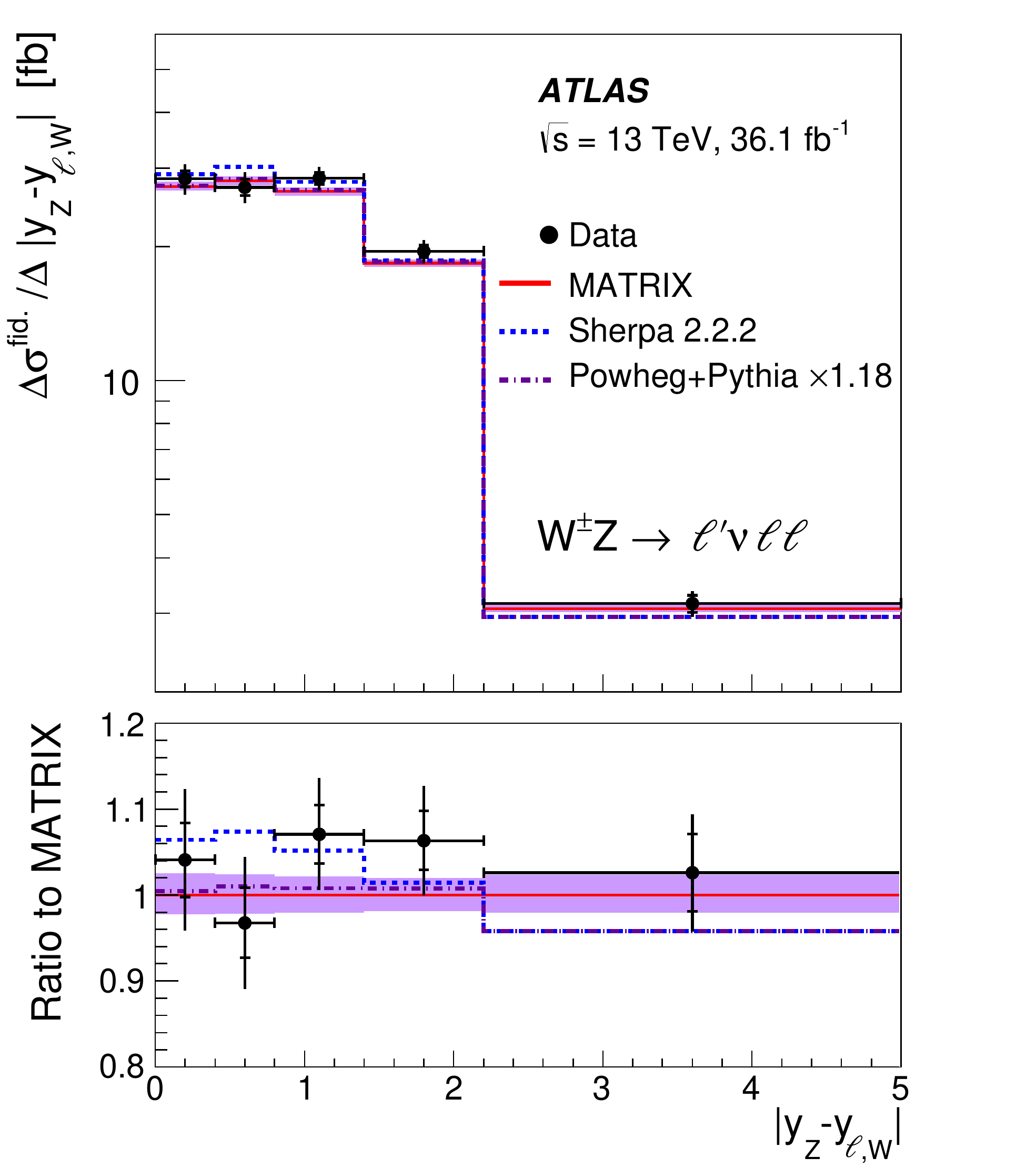}\put(-110,190){{(b)}}
\caption{
The measured $W^{\pm}Z$ differential cross section in the fiducial phase space as a function of (a) $p_{\textrm {T}}^\nu$  and (b) $|y_Z - y_{\ell,W}|$.
The inner and outer error bars on the data points represent the statistical and total uncertainties, respectively.
The measurements are compared with the NNLO prediction from \matrix (red line, see text for details).
The violet band shows how the QCD scale uncertainties affect the NNLO predictions.
The predictions from the \powhegpythia and \sherpa MC generators are also indicated by dotted-dashed and dashed lines, respectively.
In (a), the right vertical axis refers to the last cross-section point, separated from the others by vertical dashed lines, 
as this last bin is integrated up to the maximum value reached in the phase space and the cross section is not divided by the bin width.
}
\label{fig:DiffXSection:yZlW_all}
\end{center}
\end{figure}

The measured  differential cross sections in Figures~\ref{fig:DiffXSection:pTZW} and~\ref{fig:DiffXSection:yZlW_all} are compared with the predictions at NNLO in QCD from the \MATRIX computational framework.
The predictions from \matrix are corrected from Born-level leptons to dressed leptons using binned correction factors determined using \powhegpythia.
The correction factors are found to be mostly constant over the ranges of all differential distributions, with a mean value of $0.96$.
The predicted and measured cross sections are in good agreement.
The measurements are also compared with NLO MC predictions from \powhegpythia, after a rescaling of its predicted integrated fiducial cross section to the NNLO cross section, and to \sherpa~2.2.2 without rescaling its prediction.
Good agreement of the shapes of the measured distributions with the predictions of \powhegpythia and \sherpa~2.2.2 is observed.
The \dPhiWZ distribution, which is sensitive to QCD higher-order perturbative effects, is better described by \matrix than by \powhegpythia or \sherpa~2.2.2.

As shown in previous publications, the high energy tails of the \ptz~\cite{Aad:2012twa} and \mtwz~\cite{ATLASWZ_8TeV} observables are sensitive to aTGC, \ptz having the
disadvantage of being more subject to higher-order perturbative effects in QCD~\cite{Sapeta:2012} and electroweak theory~\cite{Biedermann:2017oae}.
This is seen also here with larger NNLO QCD scale uncertainties predicted by \matrix for \ptz\ than for \mtwz.
No excess of data events in the tails of these distributions is observed.

The exclusive multiplicity of jets above a \pt threshold of $25$~\GeV\ unfolded at particle level is presented in Figure~\ref{fig:unfoldResultNjetMjj}(a).
The measurements are compared with predictions from \SHERPA~2.2.2, \sherpa~2.1 and \powhegpythia.
The \SHERPA predictions provide a better description of the ratio of $0$-jet to $1$-jet event cross sections than \powhegpythia.
However, the \SHERPA~2.2.2 prediction, which models up to one parton at NLO, tends to overestimate the cross section of events with two or more jets,
while \sherpa 2.1 agrees better with data for $N_{\text{jets}}$ up to three.
Yields of events with higher jet multiplicities are described by the parton shower modelling of the \powhegpythia MC.
Finally, the measured $W^{\pm}Z$ differential cross section as a function of the invariant mass, $m_{jj}$, of the two leading  jets with $\pT > 25$~\GeV\ is presented in Figure~\ref{fig:unfoldResultNjetMjj}(b).
The measurement is better described by the \SHERPA predictions.
The production of \wz\ in association with two jets produced as a result of electroweak processes is not included in the SM predictions presented in the figure. 
In the last $m_{jj}$ bin it amounts to $17\%$ of the measured cross section, as estimated using \sherpa 2.2.2.

\begin{figure}[!htbp]
\begin{center}
\includegraphics[width=0.4\textwidth]{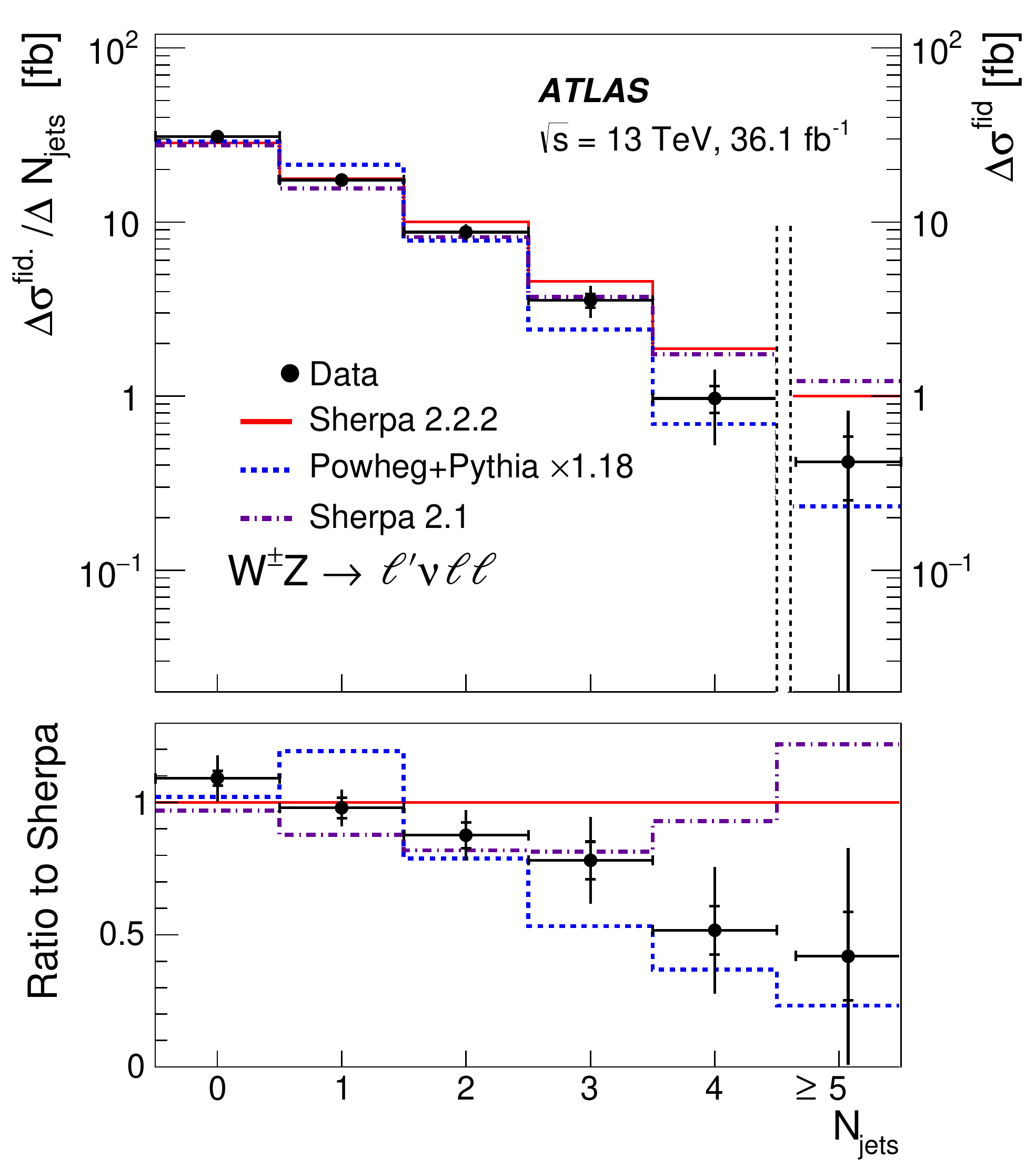}\put(-110,190){{(a)}}\quad
\includegraphics[width=0.4\textwidth]{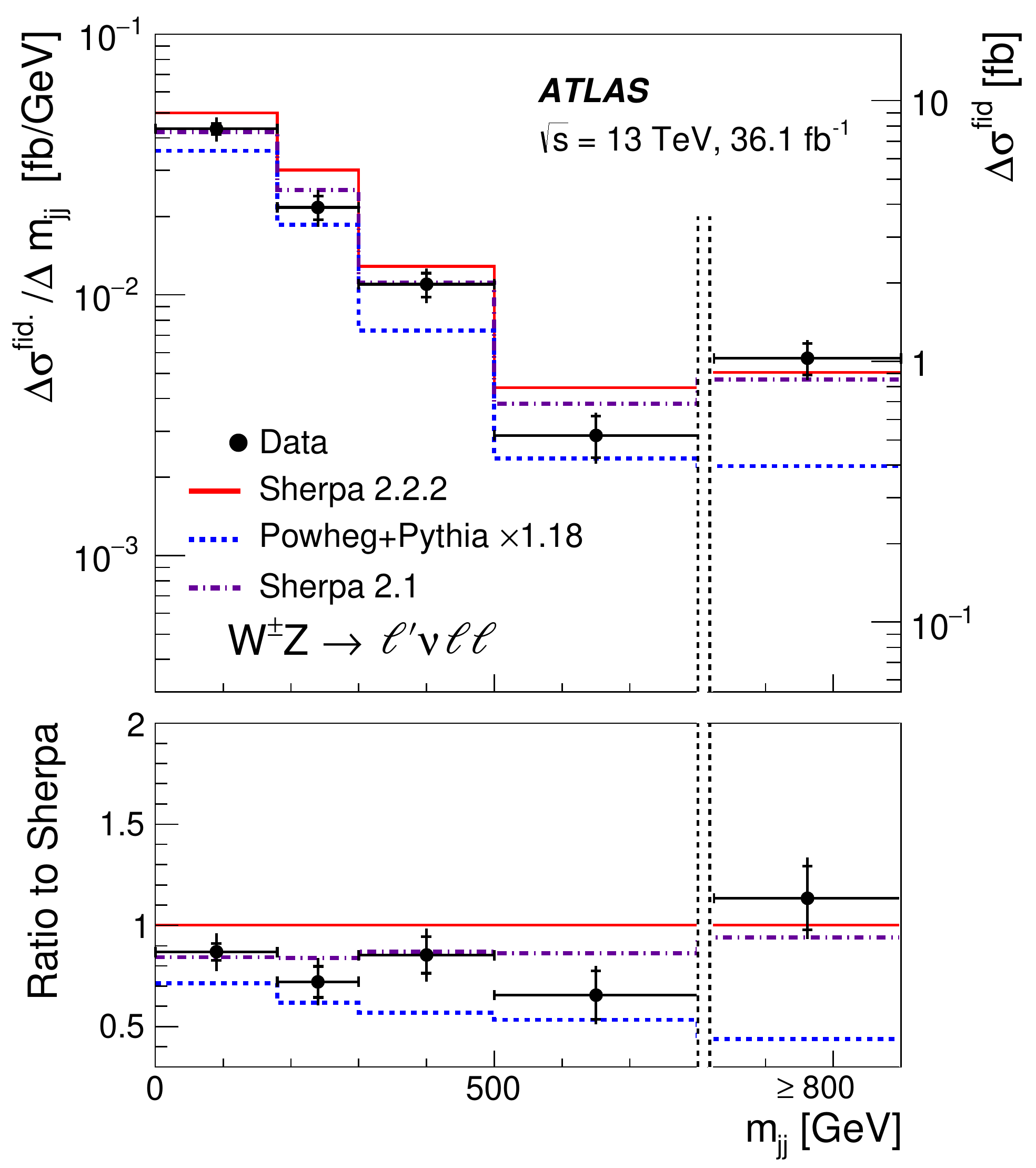}\put(-110,190){{(b)}}
\caption{The measured $W^{\pm}Z$ differential cross section in the fiducial phase space as a function of the exclusive multiplicity of jets with $p_\textrm{T} > 25$~\GeV\ (a) and of the invariant mass of the two leading jets with $p_\textrm{T} > 25$~\GeV\ (b).
The inner and outer error bars on the data points represent the statistical and total uncertainties, respectively.
The measurements are compared with the predictions from \sherpa~2.2.2 (red line), \powhegpythia (dashed blue line)  and \SHERPA~2.1 (dotted-dashed violet line).
The right vertical axis refers to the last cross-section point, separated from the others by vertical dashed lines, 
as this last bin is integrated up to the maximum value reached in the phase space and the cross section is not divided by the bin width.
}
\label{fig:unfoldResultNjetMjj}
\end{center}
\end{figure}

\newpage
\clearpage
%-------------------------------------------------------------------------------
\section{Polarisation measurement}
\label{sec:Polarisation}
\subsection{Formalism and analysis principle}\label{sec:pol_form}

The polarisation of a gauge boson  can be determined from the angular distribution of its decay products.
At the Born level, the expected angular distribution for massless fermions in the rest frame of the parent $W$ boson is given in terms of the diagonal elements \f0, \fL and \fR of the spin density matrix~\cite{Mirkes:1994eb,Groote:2012xr,Stirling:2012zt,Gounaris:1992kp} by

\begin{equation}
\label{eq:WPolarisation}
\frac{1}{\sigma_{W^{\pm}Z}}\frac{\mathrm{d}\sigma_{W^{\pm}Z}}{\mathrm{d}\cos\theta_{\ell, W}} = \frac{3}{8}f_{\text{L}}[(1\mp\cos\theta_{\ell, W})^2]+\frac{3}{8}f_{\text{R}}[(1\pm\cos\theta_{\ell, W})^2]+\frac{3}{4}f_0\sin^2\theta_{\ell, W} \; ,
 \end{equation}

where $\theta_{\ell, W}$ is defined using the helicity frame, as the decay angle of the charged lepton in the $W$ rest frame relative to the $W$ direction in the $WZ$ centre-of-mass frame, as shown in Figure~\ref{fig:AnglesForPolarisation}. The terms \f0, \fL and \fR refer to the longitudinal, transverse left-handed and transverse right-handed helicity fractions, respectively, and the normalisation is chosen such that $\f0 + \fL + \fR = 1$.
In the equation, the upper and lower signs correspond to $W^+$ and $W^-$ bosons, respectively.
All dependencies on the azimuthal angle are integrated over.

The expected angular distribution of the lepton decay products of the $Z$ boson is described by the generalisation of Equation~(\ref{eq:WPolarisation})~\cite{Mirkes:1994eb,Groote:2012xr,Stirling:2012zt}:

\begin{eqnarray}
\label{eq:ZPolarisation}
\frac{1}{\sigma_{W^{\pm}Z}}\frac{\mathrm{d}\sigma_{W^{\pm}Z}}{\mathrm{d}\cos\theta_{\ell, Z}} & = & \frac{3}{8}f_{\text{L}}(1+2 \alpha \cos\theta_{\ell, Z}+\cos^2\theta_{\ell, Z})\nonumber\\
 & + & \frac{3}{8}f_{\text{R}}(1+\cos^2\theta_{\ell, Z}-2 \alpha  \cos\theta_{\ell, Z})\nonumber\\
 & + & \frac{3}{4}f_0\sin^2\theta_{\ell, Z} \; ,
\end{eqnarray}
 
where $\theta_{\ell, Z}$ is defined using the helicity frame, as the decay angle of the negatively charged lepton in the $Z$ rest frame relative to the $Z$ direction in the $WZ$ centre-of-mass frame. 
The parameter $\alpha =(2c_vc_a)/(c_v^2+c_a^2)$ is expressed in terms of the vector $c_v=-\frac{1}{2}+2\sin^2{\theta_{\mathrm{W}}^{\mathrm{eff}}}$ and axial-vector $c_a=-\frac{1}{2}$  couplings of the $Z$ boson to leptons, respectively, where the effective value of the Weinberg angle $\sin^2{\theta_{\mathrm{W}}^{\mathrm{eff}}}=0.23152$~\cite{PhysRevD.98.030001} is used.
Equation~(\ref{eq:ZPolarisation}) also holds for the contribution from $\gamma^*$ and its interference with the $Z$ boson, with appropriate $c_v$ and $c_a$ coefficients.
The tight invariant mass window of $\pm 10$~\GeV\ around the nominal $Z$ boson mass minimises the contribution from $\gamma^*$, although all the helicity fractions presented here are effective fractions, containing the small contribution from $\gamma^*$.

\begin{figure}[!htbp]
\begin{center}
\includegraphics[width=0.38\textwidth]{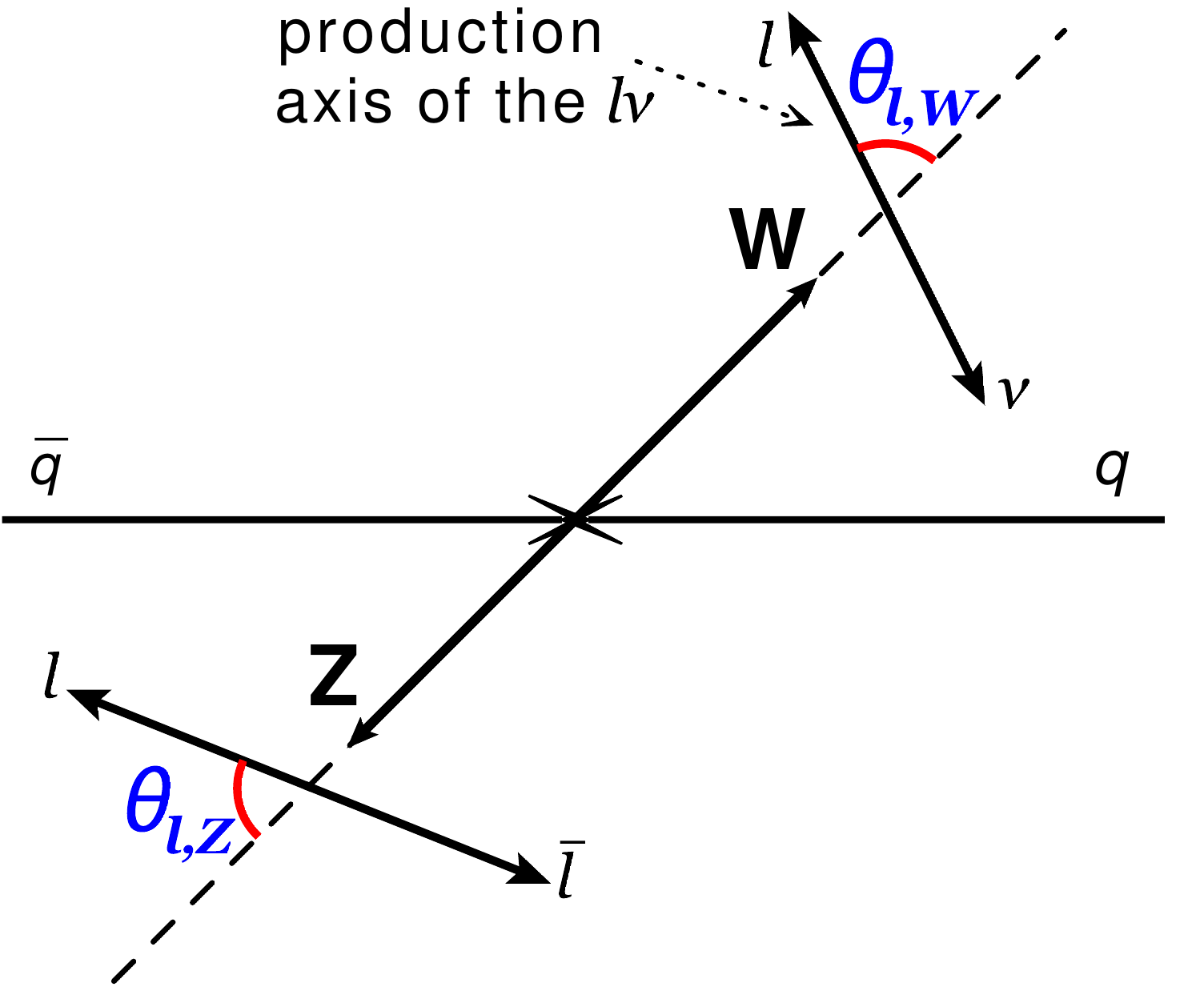}
\caption{The decay angle $\theta_{\ell, W(Z)}$ is defined as the angle between the negatively (positively for $W^+$) charged lepton produced in the decay of the $W$ ($Z$) 
boson as seen in the $W$ ($Z$) rest frame and the direction of the $W$ ($Z$) which is given in the $WZ$ centre-of-mass frame.}
\label{fig:AnglesForPolarisation}
\end{center}
\end{figure}

Equations~(\ref{eq:WPolarisation}) and~(\ref{eq:ZPolarisation}) are valid only when the full phase space of the leptonic decays of the  gauge bosons is accessible.
Restrictions on the \pt\ and $\eta$ values of the charged decay lepton or of the neutrino suppress events at $\left| \cos\theta_{\ell, W(Z)} \right| \sim 1$, as shown in Figure~\ref{fig:Polarisation:fidPS_pol_dist}, and the analytic expressions of Equations (\ref{eq:WPolarisation}) and (\ref{eq:ZPolarisation}) cannot be used to extract the helicity fractions.
Simulated templates therefore must be used.

Another major difficulty arises for the $W$ boson from incomplete knowledge of the neutrino momentum.
The large angular coverage of the ATLAS detector enables measurement of the missing transverse momentum, which can be identified as the transverse momentum
of the neutrino. 
The neutrino longitudinal momentum $p_z^\nu$ is obtained using the $W$ mass constraint. 
Solving the corresponding equation leads to a twofold ambiguity, which is resolved by choosing
the solution with the smaller $|p_z^\nu|$.
If the measured transverse mass is larger than the
nominal $W$ mass, no real solutions exist for $p_z^\nu$.
The most likely cause is that the measured $\met$ is larger than the actual neutrino \pT. 
In this case, the best estimate is obtained by choosing the real part of the complex solutions. 
As an alternative to the $\cos\theta_{\ell,W}$ observable using this reconstruction of the neutrino momentum, a ``transverse helicity'' observable introduced in Ref.~\cite{ATLAS:2012au} was tested, but a similar or lower sensitivity for the measurement of the \f0 helicity fraction for $W$ bosons was obtained, so it was not pursued further.

\begin{figure}[!htbp]
\begin{center}
\includegraphics[width=.38\textwidth]{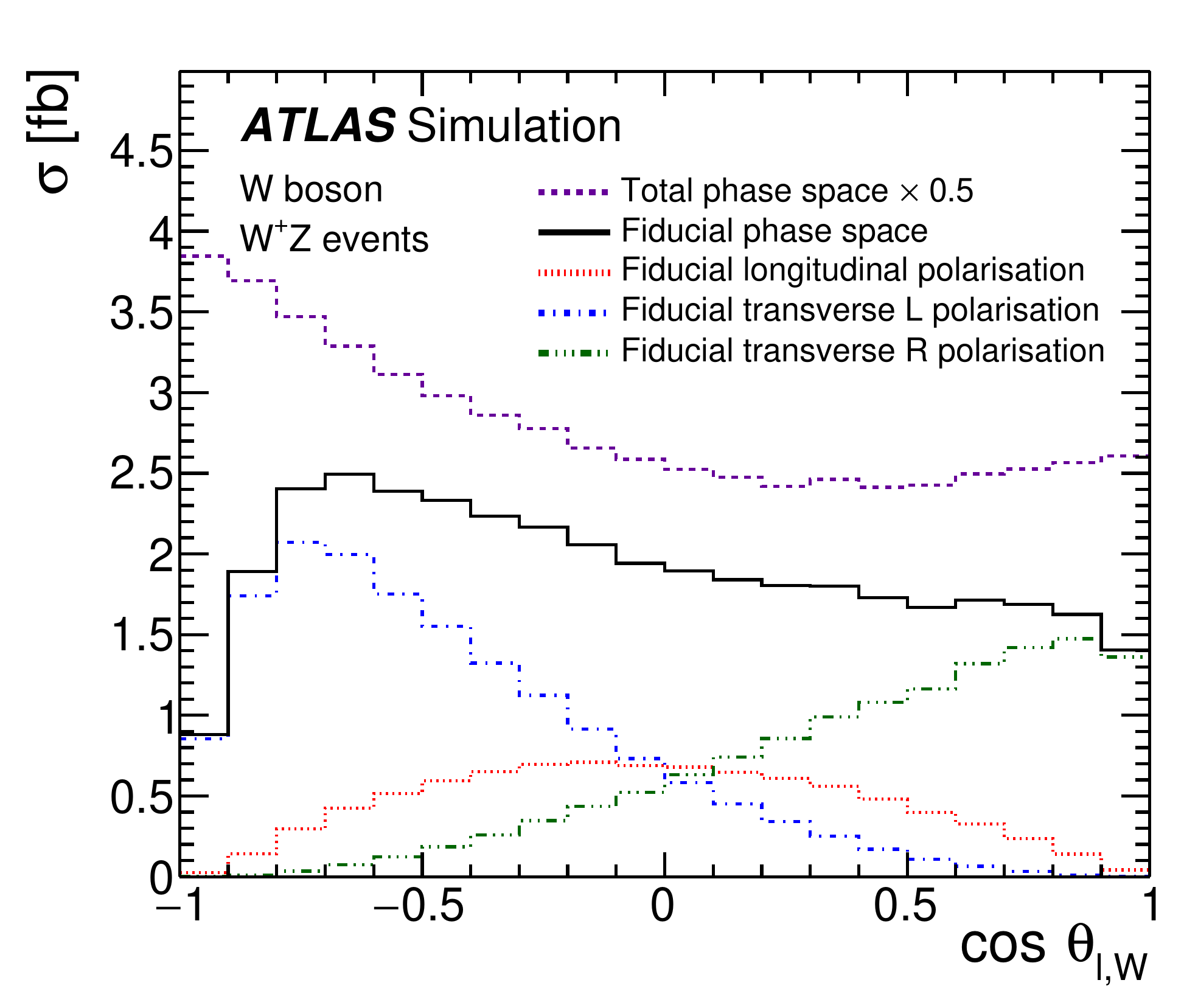}\put(-25,125){{(a)}}
\includegraphics[width=.38\textwidth]{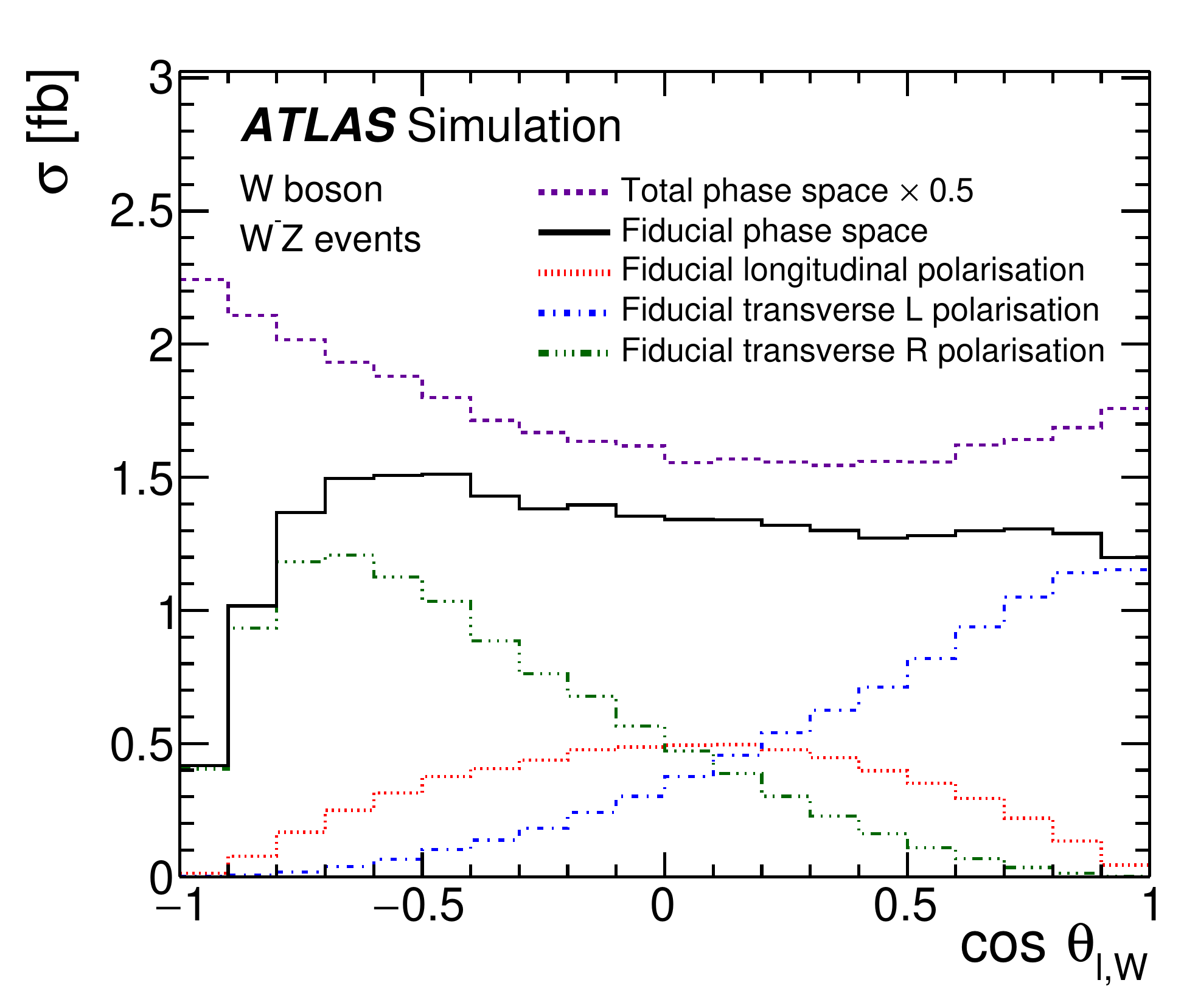}\put(-25,125){{(b)}}\\
\includegraphics[width=.38\textwidth]{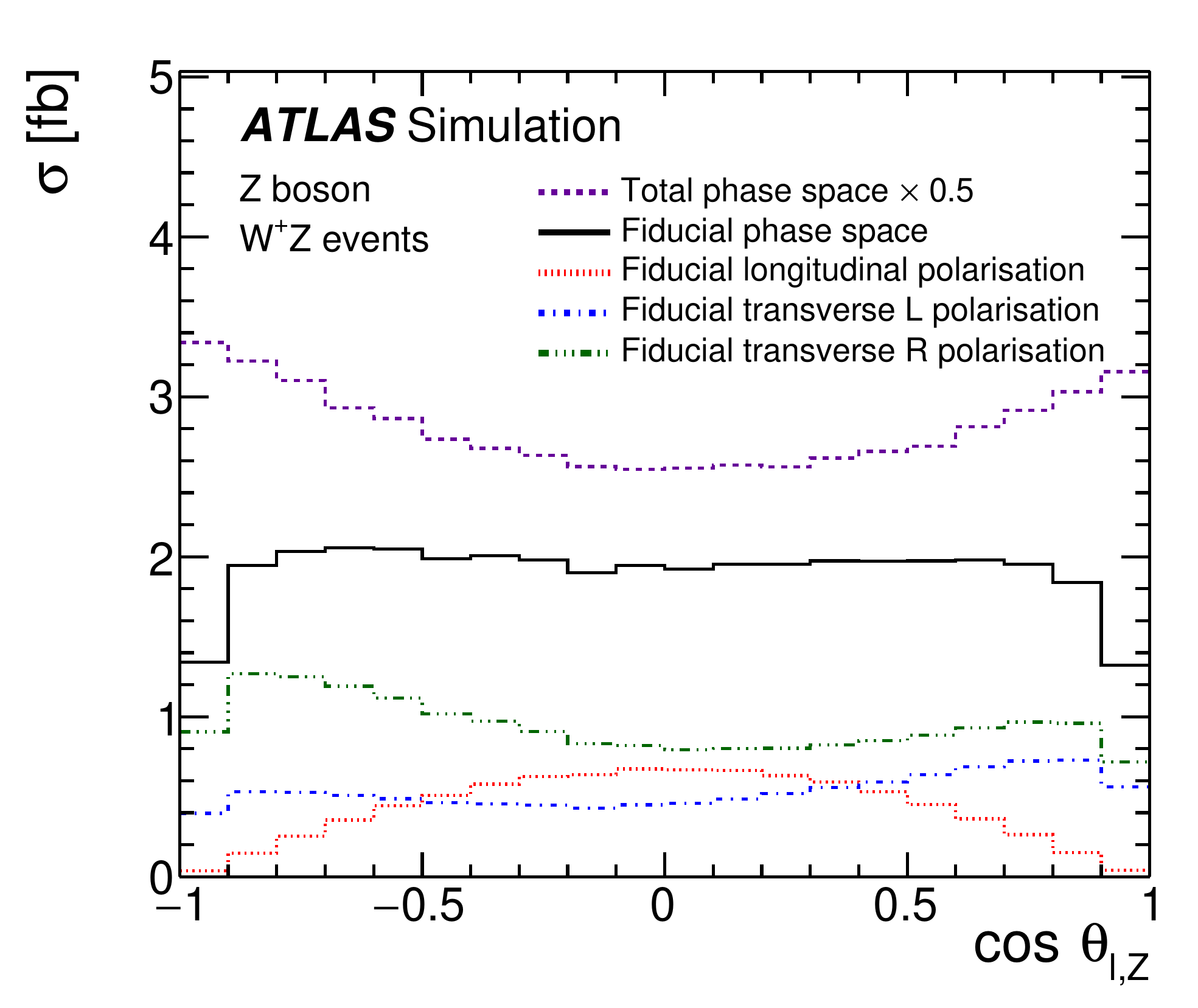}\put(-25,125){{(c)}}
\includegraphics[width=.38\textwidth]{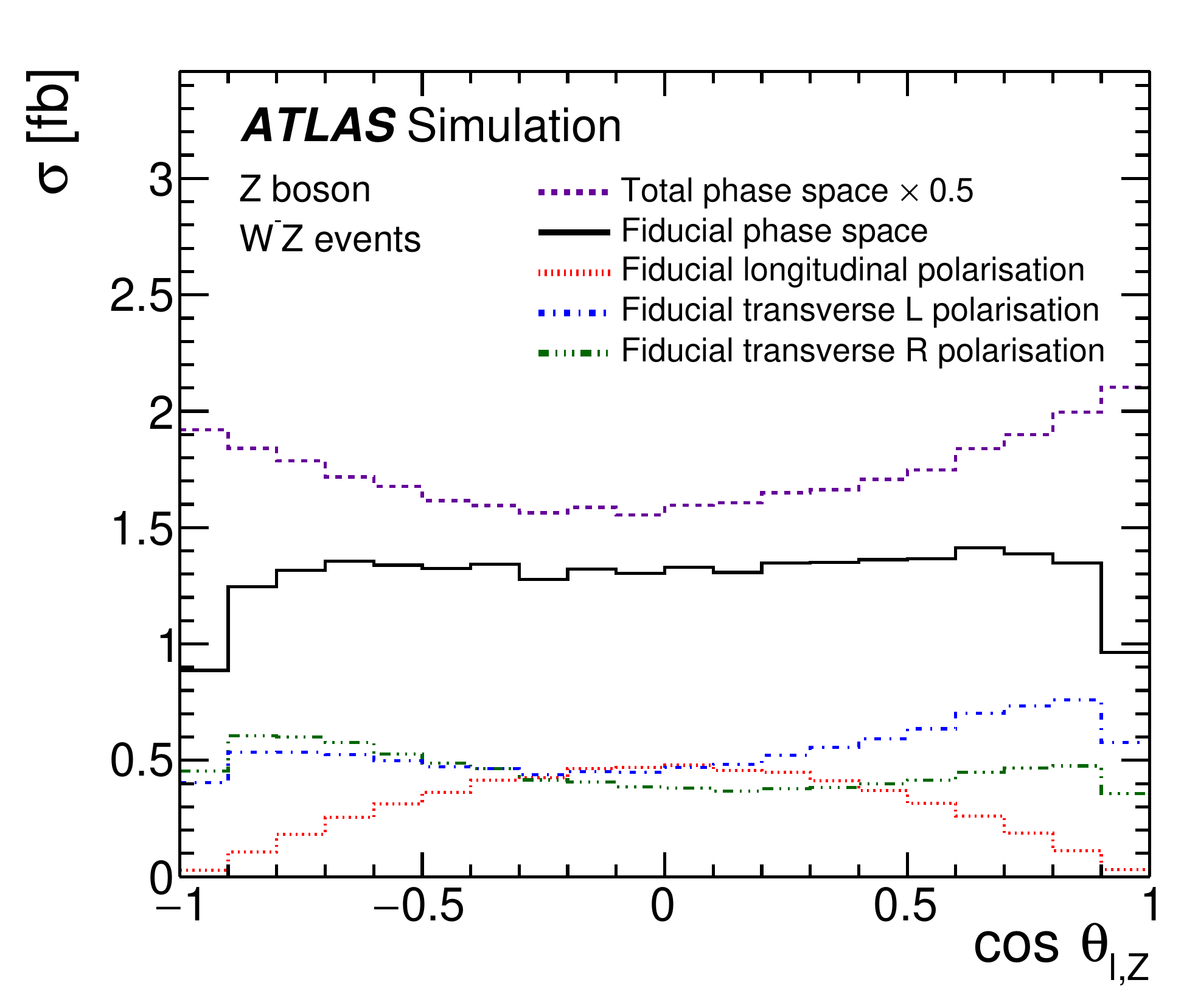}\put(-25,125){{(d)}}
\caption{Distributions in the total and fiducial phase space at particle level of the variables (a, b) $\cos \theta_{\ell, W}$ and   (c,d) $\cos \theta_{\ell, Z}$ for (a, c) $W^+ Z$ and (b, d)  $W^- Z$events.
The black line corresponds to the sum of all helicity states. 
The red, blue and green lines correspond to the purely longitudinal, transverse left-handed and transverse right-handed helicity components, respectively.
The distributions are obtained using the \powhegpythia MC.
All four decay channels, $eee$, $e\mu\mu$, $\mu ee$, and $\mu\mu\mu$, are added together.
 }
\label{fig:Polarisation:fidPS_pol_dist}
\end{center}
\end{figure}

\medskip

For the polarisation  measurements, all four decay channels, $eee$, $e\mu\mu$, $\mu ee$, and $\mu\mu\mu$, are added together.
The measurements of $W$ and $Z$ boson polarisation are performed separately for $W^+Z$, $W^-Z$ and \wz\ events.
To allow the datasets of both $W$ boson charges to be combined for the measurement in \wz events, $\cos\theta_{\ell,W}$ is multiplied by the sign
of the lepton charge $q_\ell$.
Figures~\ref{fig:Polarisation:pol_dist}(a),~(b) present the reconstructed distributions for \wz\ events of $q_\ell \cdot \cos\theta_{\ell,W}$ for the $W$ bosons  and of $\cos\theta_{\ell,Z}$ for $Z$ bosons.
The MC predictions provide a good description of the shapes of the data distributions.

The helicity parameters \f0 and \fLR are measured in \wz\ events separately for $W$ and $Z$ bosons using a binned profile-likelihood fit~\cite{frequentist} of templates of the three helicity states to the $q_\ell \cdot \cos\theta_{\ell,W}$ and $\cos\theta_{\ell,Z}$  distributions.
The equation $\f0 + \fR + \fL = 1$ is used to constrain the independent parameters of the fit to \f0, \fLR and the integrated fiducial cross section.
The templates of $q_\ell \cdot \cos\theta_{\ell,W}$ and $\cos\theta_{\ell,Z}$ distributions for each of the three helicity states of the $W$ and $Z$ bosons are extracted from the \powhegpythia MC sample~\cite{ATLAS:2012au}.
For each of the gauge bosons, generically denoted as $V$, the predicted helicity fractions of \powhegpythia MC events are determined as a function of $p_{\mathrm{T}}^V$ and $y_V$ by fitting the analytic functions of equations (\ref{eq:WPolarisation}) and (\ref{eq:ZPolarisation}) to the predicted $\cos\theta_{\ell,V}$ distributions in the total phase space.
Two dimensional bins as a function of $p_{\mathrm{T}}^V$ and $y_V$ are used. The bin boundaries are optimised such that possible bias on the evolution of the extracted helicity fractions is minimised.
The MC templates at detector level  representing longitudinal, left- and right-handed states of the $W$ boson are then obtained by reweighting of \powhegpythia MC events according to

\begin{equation*}\label{eq:weightsWpol}
\frac{ \left . {\frac{1}{ \sigma_{W^{\pm}Z} } \frac{\mathrm{d}\sigma_{W^{\pm}Z} } {\mathrm{d}\cos\theta_{\ell, W} } }\right|_{\mathrm{L}/\mathrm{0}/\mathrm{R}} } {\frac{3}{8}f_{\mathrm{L}}^{\mathrm{gen.}}(1\mp\cos\theta_{\ell, W})^2 + \frac{3}{8}f_{\mathrm{R}}^{\mathrm{gen.}} (1\pm\cos\theta_{\ell, W})^2 + \frac{3}{4}f_\mathrm{0}^{\mathrm{gen.}}\sin^2\theta_{\ell, W} }\; ,
\end{equation*}

where

\begin{equation*}
\left . \frac{1}{\sigma_{W^{\pm}Z}}\frac{\mathrm{d}\sigma_{W^{\pm}Z}}{\mathrm{d}\cos\theta_{\ell, W}} 
\right |
   \begin{array}{r}
      \mathrm{L}\\
      \mathrm{0} \\
      \mathrm{R}
   \end{array}
   = 
 \frac{3}{8} \left \{
   \begin{array}{r}
      (1\mp\cos\theta_{\ell, W})^2 \\
      2 \sin^2\theta_{\ell, W}\\
      (1\pm\cos\theta_{\ell, W})^2
   \end{array}
\right . \, ,
\end{equation*}

and where $f_{\mathrm{L/0/R}}^{\mathrm{gen.}}$ are the  helicity fractions at generator level, extracted by the fit, of \powhegpythia MC events.
Similar equations hold for the polarisation of the $Z$ boson. 
The procedural uncertainty of the reweighting method for the generation of MC templates was estimated to be below $0.5$\%.
Helicity fractions are extracted by the template fit at detector level.
To be expressed in a fiducial phase space at particle level, each helicity fraction is then corrected independently for detector efficiencies and QED final-state radiation effects using factors obtained from the simulation.
Measured helicity fractions are thus reported at particle level for a fiducial phase space which follows the definition of Section~\ref{sec:phase_space} with the difference that leptons with kinematics defined before QED final state radiation (``Born leptons'') are used instead of dressed leptons.
Experimental systematic uncertainties detailed in Section~\ref{sec:systematics} are considered and treated as nuisance parameters with an assumed Gaussian distribution.  
Theoretical systematic uncertainties due to the modelling in the event generator used to evaluate the helicity templates are considered.
The effects of PDF and QCD scale uncertainties are estimated as detailed in Section~\ref{sec:systematics}.
An additional modelling uncertainty is considered and estimated by comparing predictions from the \powhegpythia and \mcatnlo MC event generators for the shape of helicity template distributions.

\begin{figure}[!htbp]
\begin{center}
\includegraphics[width=.4\textwidth]{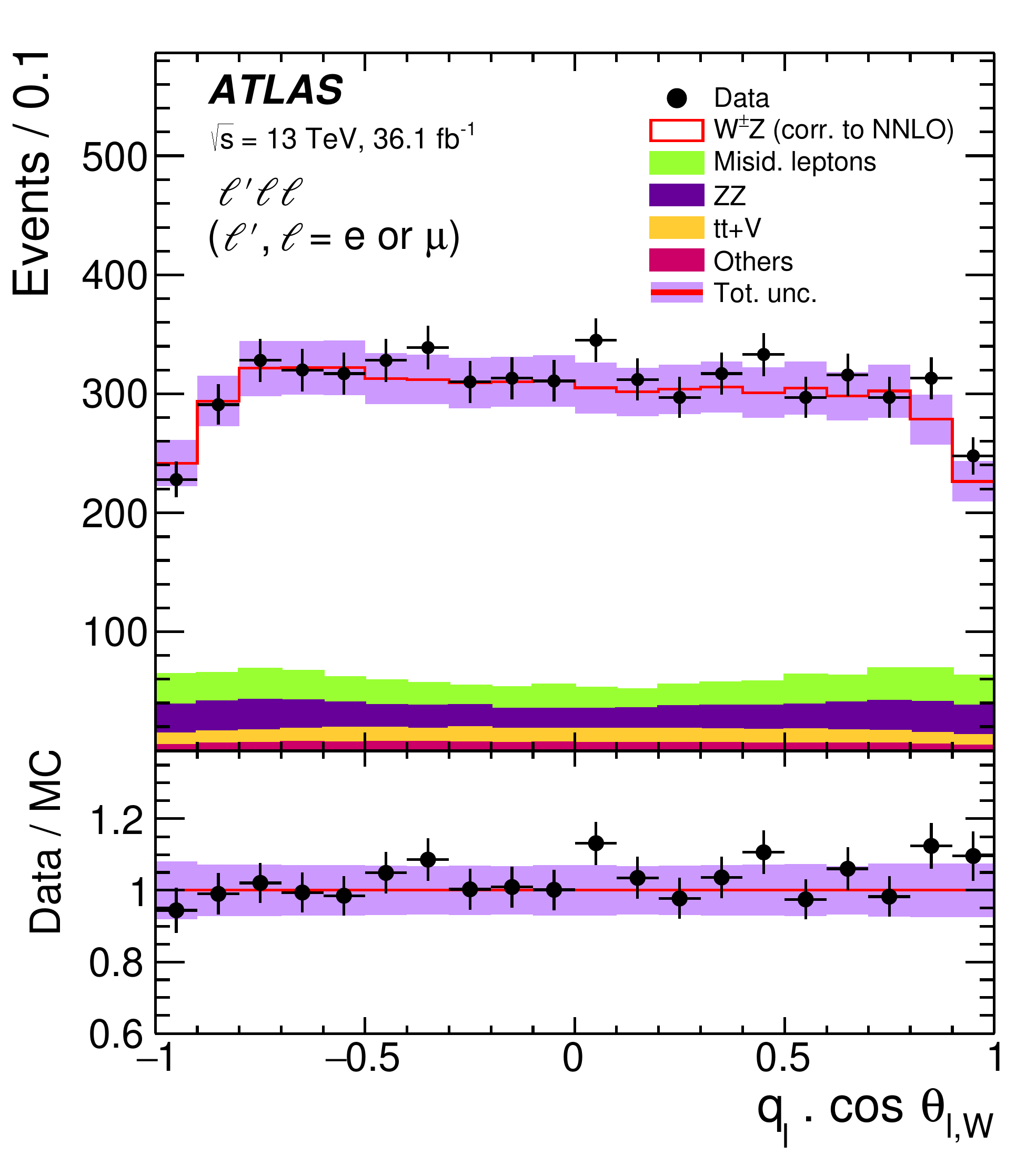}\put(-30,100){{(a)}}
\includegraphics[width=.4\textwidth]{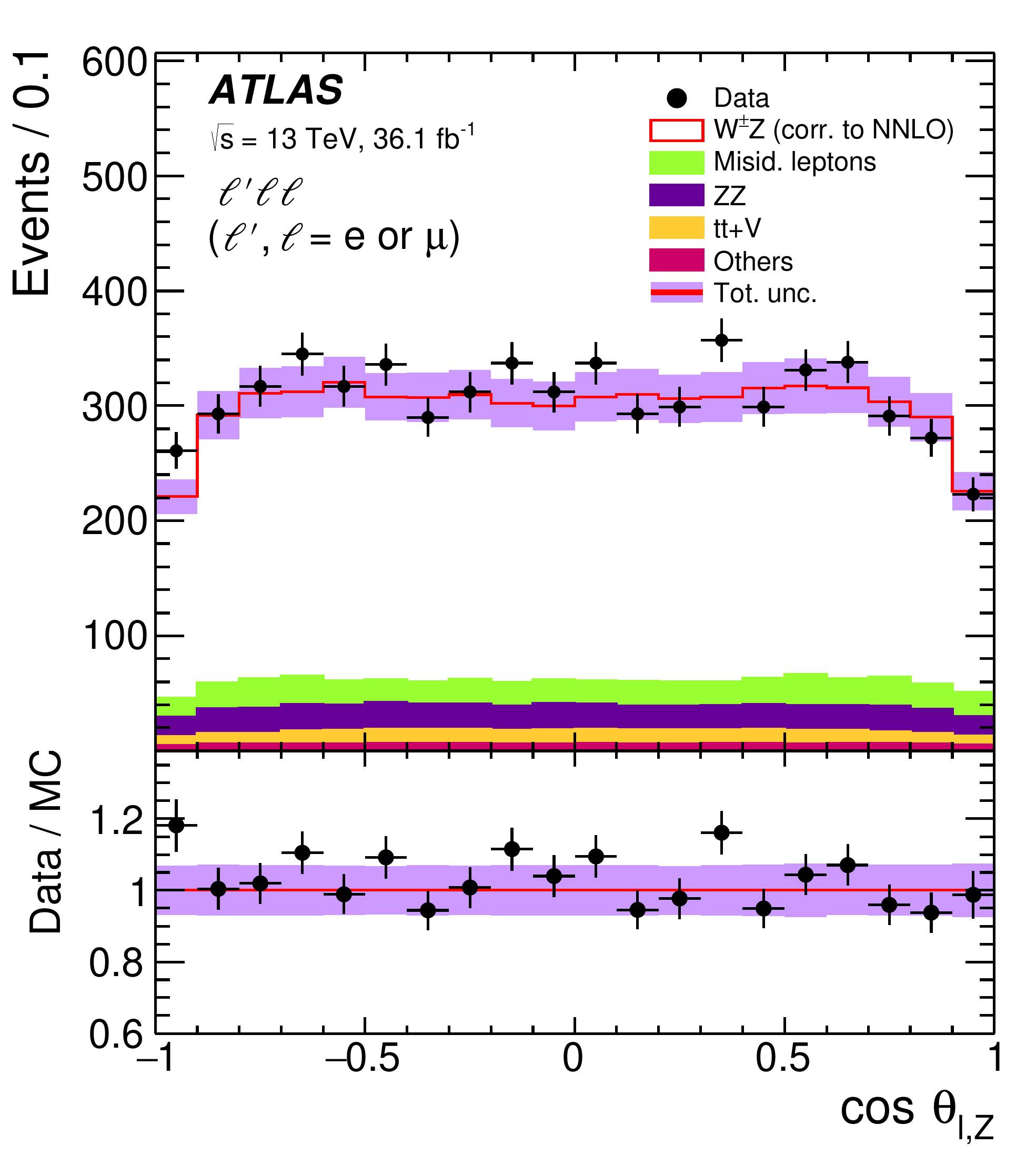}\put(-30,100){{(b)}}
\caption{The detector-level distributions for the sum of all channels of the variables (a)  $q_\ell \cdot \cos \theta_{\ell, W}$ and   (b) $\cos \theta_{\ell, Z}$.
The points correspond to the data with the error bars representing the statistical uncertainties, and the histograms correspond to the predictions of the different SM processes. 
The sum of the background processes with misidentified leptons is labelled ``Misid. leptons''.
The \powhegpythia MC prediction is used for the \wz\ signal contribution. 
It is scaled by a global factor of $1.18$ to match the \NNLO cross section predicted by \matrix.
The open red histogram shows the total prediction; the shaded violet band
is the total uncertainty of this prediction.
The lower panels in each figure show the ratio of the data points to the
open red histogram with their respective uncertainties.
}
\label{fig:Polarisation:pol_dist}
\end{center}
\end{figure}

\subsection{Results}\label{sec:pol_results}

The measurements of \f0 and \fLR are summarised in Table~\ref{tab:Polarisation:PolRes}, where they are compared with the predictions from \powhegpythia.
The \powhegpythia MC sample was generated at LO in the electroweak formalism using the $G_\mu$ scheme with $\sin^2{\theta_{\mathrm{W}}}=0.2229$.
This choice impacts the predicted \fLR values which depend on the chosen value of the Weinberg angle via the angular coefficient $A_4$~\cite{Collins:1977iv,Mirkes:1994eb}.
The impact of the value of $\sin^2{\theta_{\mathrm{W}}}$ on \fLR is estimated using  \MCFM~\cite{Campbell:1999ah,Campbell:2011bn,Campbell:2015qma} calculations with two electroweak schemes, the $G_\mu$ scheme and a scheme where the value $\sin^2{\theta_{\mathrm{W}}^{\mathrm{eff}}}=0.23152$ is imposed.
The difference between the two calculations is used to correct the \fLR values predicted by \powhegpythia.

\begin{table}[!htbp]
\caption{
Measured helicity fractions in the fiducial phase space with Born-level leptons, for \wz\, $W^+Z$ and $W^-Z$ events.
The total uncertainties in the measurements are reported.
The measurements are compared with predictions at electroweak LO from \powhegpythia and \matrix corrected to $\sin^2{\theta_{\mathrm{W}}^{\mathrm{eff}}}=0.23152$.
The uncertainties on the \powhegpythia prediction include QCD scale and PDF uncertainties; the uncertainties in the \matrix prediction include QCD scale uncertainties.
}
\label{tab:Polarisation:PolRes}
\begin{center}
\resizebox{0.97\textwidth}{!}{
\begin{tabular}{
 l | 
 S@{\,}@{$\pm$}S
 S@{\,}@{$\pm$}S  
 S@{\,}@{$\pm$}S |
 S@{\,}@{$\pm$}S 
  S@{\,}@{$\pm$}S 
  S@{\,}@{$\pm$}S 
  }
\toprule
                                   &  \multicolumn{6}{c}{\f0}                                                                                                                 & \multicolumn{6}{c}{\fLR} \\ 
                                   & \multicolumn{2}{c}{Data} & \multicolumn{2}{c}{\powhegpythia}   & \multicolumn{2}{c}{\matrix}   & \multicolumn{2}{c}{Data} & \multicolumn{2}{c}{\powhegpythia}  & \multicolumn{2}{c}{\matrix} \\  
\midrule
$W^{+}$ in $W^{+}Z$ & 0.26 & 0.08                     &  0.233& 0.004                                   &0.2448 & 0.0010                    & -0.02 & 0.04                    & 0.091&0.004                                    & 0.0868 & 0.0014 \\ 
$W^{-}$ in $W^{-}Z$   & 0.32 & 0.09                     &  0.245 & 0.005                                  &0.2651 & 0.0015                    & -0.05 & 0.05                    & -0.063 & 0.006                                 & -0.034 & 0.004 \\ 
$W^{\pm}$ in $W^{\pm}Z$ & 0.26 & 0.06              &  0.2376 & 0.0031                             &0.2506 & 0.0006                    & -0.024 & 0.033                 & 0.0289 & 0.0022                              & 0.0375 & 0.0011    \\ 
$Z$ in $W^{+}Z$                & 0.27 & 0.05              &  0.225 & 0.004                                  &0.2401 & 0.0014                   & -0.32 & 0.21                     & -0.297 & 0.021                                 & -0.262 & 0.009     \\ 
$Z$ in $W^{-}Z$                 & 0.21& 0.06               &  0.235 & 0.005                                  &0.2389 & 0.0015                   & -0.46 & 0.25                     & 0.052 & 0.023                                  & 0.0468 & 0.0034       \\ 
$Z$ in $W^{\pm}Z$            & 0.24 & 0.04              &  0.2294 & 0.0033                              &0.2398 & 0.0014                    & -0.39 & 0.16                    & -0.156 & 0.016                                 & -0.135 & 0.006   \\ 
\bottomrule
\end{tabular}
  }

\end{center}
\end{table}

\begin{table}[!htbp]
\caption{Summary of the absolute uncertainties in the helicity fractions \f0 and \fLR measured in \wz\ events for $W$ and $Z$ bosons.
}
\label{tab:SystematicSummary_pol}
\begin{center}
%------ in absolute values
\begin{center}
{\footnotesize 
\begin{tabular}{ l r  r r r}
\toprule 
                 &     \multicolumn{2}{c}{$W^{\pm}$ in $W^{\pm}Z$}                     &               \multicolumn{2}{c}{$Z$ in $W^{\pm}Z$}                                   \\  
                 &      \f0                     & \fLR        &        \f0                 & \fLR                       \\  
%\midrule 
% \multicolumn{5}{c}{Absolute Uncertainties}  \\ 
\midrule 
 $e$ energy scale and id. efficiency  & 0.0024& 0.0004  & 0.005& 0.0021\\  
 $\mu$ momentum scale and id. efficiency      & 0.0013& 0.0027  & 0.0018& 0.008\\  
 $E_{\mathrm{T}}^{\mathrm{miss}}$ and jets       & 0.0024& 0.0010  & 0.0017& 0.005\\  
 Pile-up    & 0.005& 0.00009   & 0.0014& 0.005\\  
 Misid. lepton background   & 0.031& $<$ 0.001  & 0.007 & 0.019\\  
 $ZZ$ background& 0.009& 0.0004  & 0.0007& 0.0012\\  
 Other backgrounds  & 0.0012& 0.0005  & 0.0018 & 0.005 \\  
\midrule 
 QCD scale  & 0.0008& 0.0013  & 0.0004& 0.008\\  
 PDF        & 0.0011& 0.0009  & 0.00004 & $<$ 0.00001\\  
 Modelling  & 0.004 & 0.007  & 0.0015 & 0.0028\\  
\midrule
 Total systematic uncertainty & 0.033 & 0.008  & 0.009 & 0.024\\  
 Luminosity       & 0.0015 & $<$ 0.0001  & $<$ 0.0001& 0.0008\\  
Statistics    & 0.06 & 0.032  & 0.04 & 0.15 \\  
\midrule
Total     & 0.06 & 0.033  & 0.04 & 0.16\\  
\bottomrule
\end{tabular}
  }
\end{center}

\end{center}
\end{table}

The longitudinal helicity fraction  \f0 of the $Z$ boson is measured with an observed significance of $6.5 \,\sigma$, compared to $6.1 \, \sigma$ expected.
The longitudinal helicity fraction  \f0 of the $W$ boson is more difficult to extract than for the $Z$ boson and has a larger uncertainty.
This measurement establishes the presence of longitudinally polarised $W$ bosons with an observed significance of $4.2 \,\sigma$, compared to $3.8 \, \sigma$ expected.
Table~\ref{tab:SystematicSummary_pol} shows the main sources of uncertainty in the measurement of the helicity fractions.
The measurements are dominated by statistical uncertainties. 
Good agreement of the measured helicity fractions of both the $W$ and $Z$ bosons with the predictions from \powhegpythia and \matrix is observed.
Measured \f0 values agree within $1 \sigma$ with the prediction, while \fLR values agree within $2 \, \sigma$.
The \powhegpythia and \matrix predictions are only at NLO and NNLO in QCD, respectively, but, more importantly for polarisation, 
both calculations use only LO electroweak matrix elements.
Therefore, and also because of the still large statistical uncertainties in the measurements, no stringent constraints nor clear inconsistencies between measurements and predictions can be deduced.
The values of \f0 and \fLR measured in \wz\ events are shown in Figure~\ref{fig:Polarisation:pol_Wpm} for the $W$ and $Z$ bosons, respectively.

\begin{figure}[!htbp]
\begin{center}
\includegraphics[width=.4\textwidth]{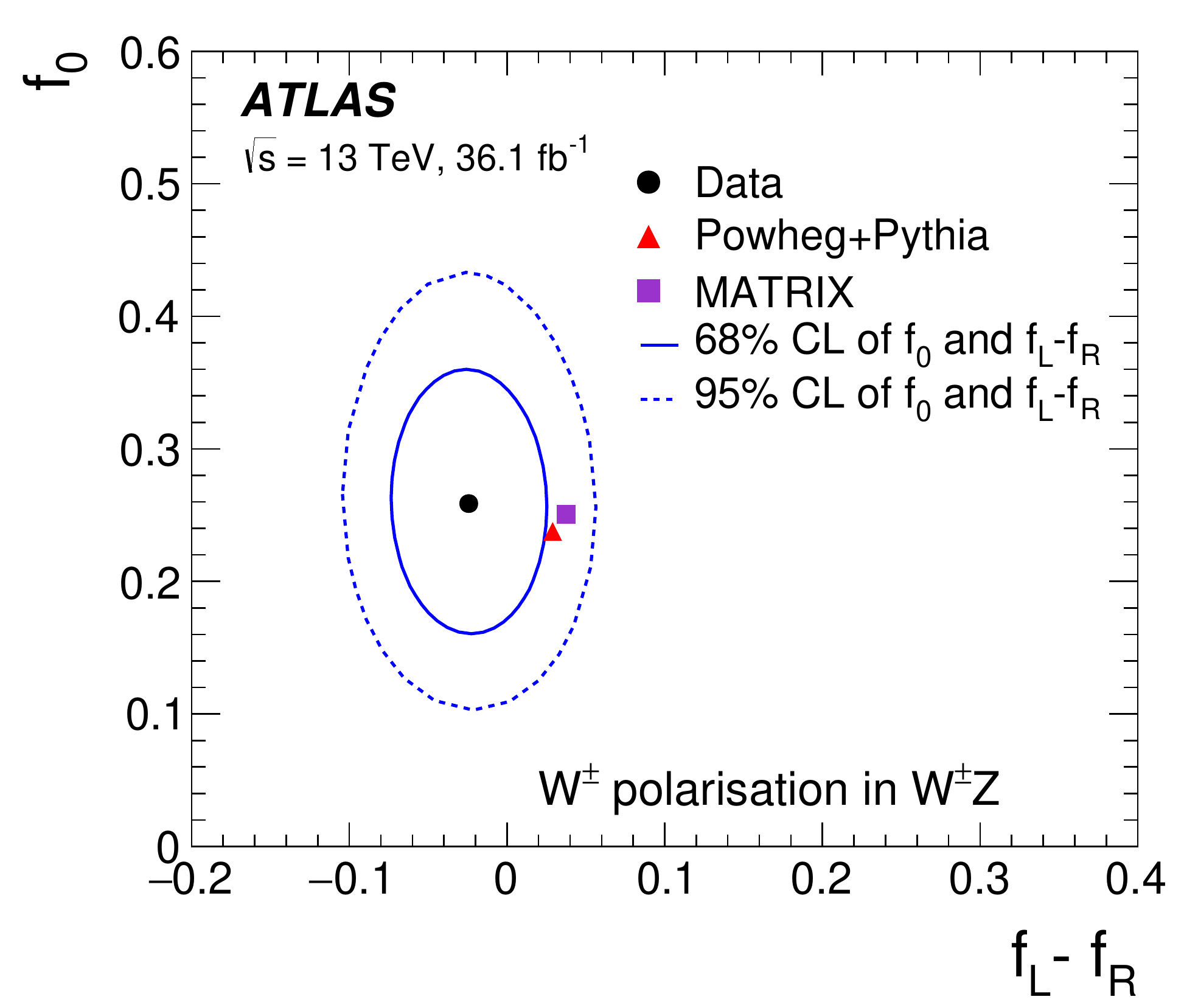}\put(-30,70){{(a)}}
\includegraphics[width=.4\textwidth]{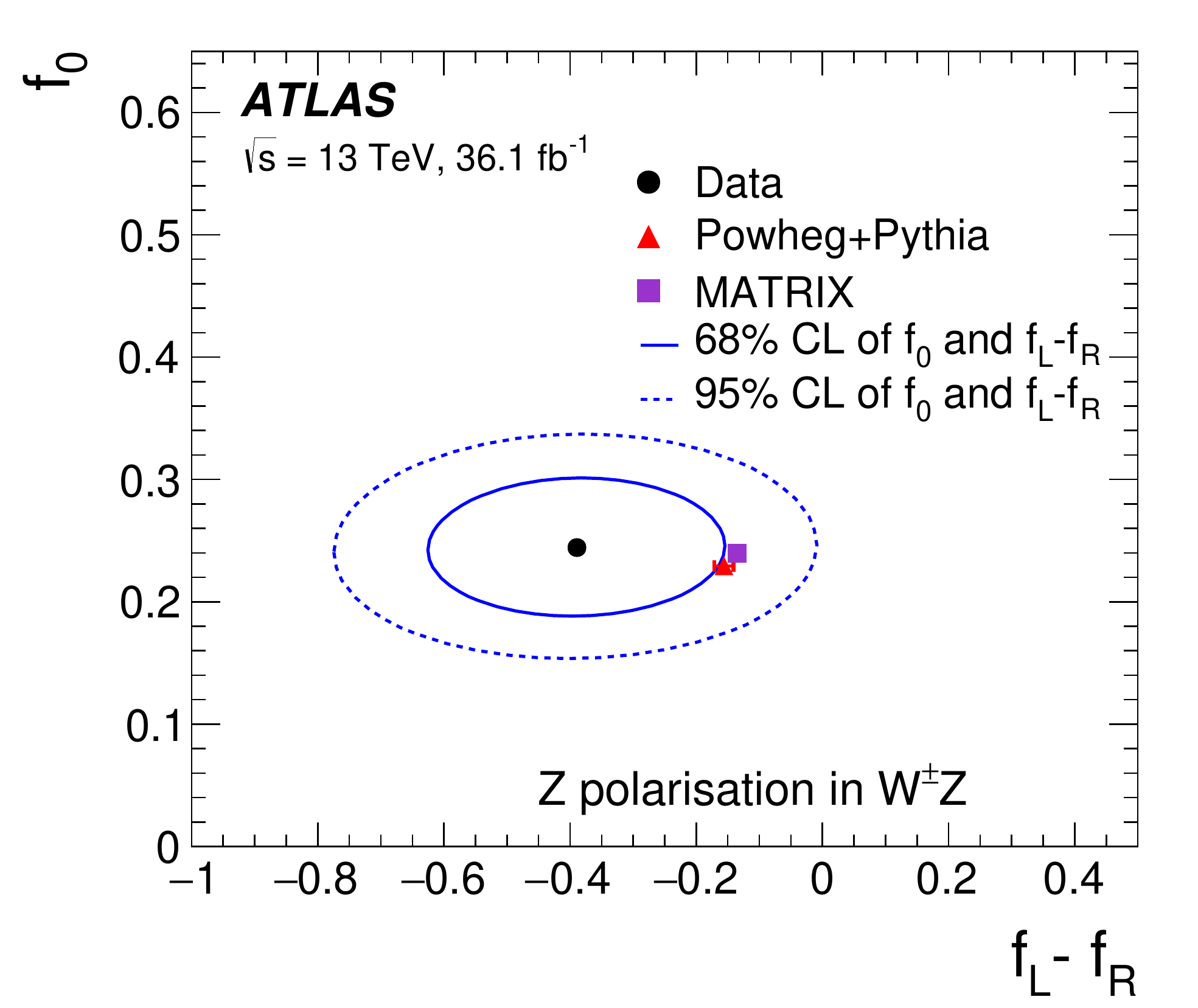}\put(-30,70){{(b)}}\caption{ Measured helicity fractions \f0 and \fLR for (a) the $W$ and (b) $Z$  bosons in $W^{\pm}Z$ events, compared with predictions at LO for the electroweak interaction and with $\sin^2{\theta_{\mathrm{W}}}=0.23152$ from \powhegpythia (red triangle) and \matrix (purple square).
The effect of PDF and QCD scale uncertainties on the \powhegpythia prediction and the effect of QCD scale uncertainties on the \matrix prediction are of the same size as the triangle marker.
The full and dashed ellipses around the data points correspond to one and two standard deviations, respectively.
 }
\label{fig:Polarisation:pol_Wpm}
\end{center}
\end{figure}

\FloatBarrier

%-------------------------------------------------------------------------------
\section{Conclusion}
\label{sec:conclusion}
Measurements of \wz\ production cross sections in $\sqrt{s} = 13$~\TeV\ $pp$ collisions at the LHC are presented.
The data analysed were collected with the ATLAS detector in $2015$ and $2016$ and correspond to an integrated luminosity of $36.1~\ifb$.
The measurements use leptonic decay modes of the gauge bosons to electrons or muons and are performed
in a fiducial phase space closely matching the detector acceptance.
The measured inclusive cross section in the fiducial region for leptonic decay modes (electrons or muons) is
$\sigma_{W^\pm Z \rightarrow \ell^{'} \nu \ell \ell}^{\textrm{fid.}} =  63.7  \, \pm~1.0~\textrm{(stat.)} \, \pm~2.3~\textrm{(syst.)} \, \pm~1.4~\textrm{(lumi.)}$~fb,
in agreement with the NNLO Standard Model expectation of $61.5^{+1.4}_{-1.3}$~fb.
The ratio of the cross sections for $W^+Z$ and $W^-Z$ production is also measured.
The \wz\ production cross section is measured as a function of several kinematic variables and compared with SM predictions at NNLO from the \matrix calculation and at NLO from the \powhegpythia and \SHERPA MC event generators.
The differential cross-section distributions are fairly well described by the theory predictions, with the exception of the jet multiplicity.  The \MATRIX calculations show the best agreement with the data.

Furthermore, an analysis of angular distributions of leptons from decays of $W$ and $Z$ bosons has been performed.
Helicity fractions  of pair-produced vector bosons are measured for the first time in hadronic collisions.
Integrated over the fiducial region, the longitudinal polarisation fractions of the $W$ and $Z$ bosons in \wz\ events are measured to be $f_0^W = 0.26 \pm 0.06$ and $f_0^Z = 0.24 \pm 0.04$, respectively, in agreement with the SM predictions at NLO in QCD and at LO for electroweak corrections, of $0.238 \pm 0.003$ and $0.230 \pm 0.003$, respectively.
The differences of the left and right transverse polarisations are also measured.% 
The measured values agree with the SM predictions within less than one and two standard deviations of their uncertainties for \f0 and \fLR, respectively. 

These polarisation measurements represent a step towards further new constraints on the electroweak symmetry 
breaking mechanism of the Standard Model, in particular by polarisation measurements in vector boson scattering.

%-------------------------------------------------------------------------------

%-------------------------------------------------------------------------------
\section*{Acknowledgements}
%-------------------------------------------------------------------------------

% Acknowledgements for papers with collision data
% Version 24-Oct-2018

% Standard acknowledgements start here
%----------------------------------------------
We thank CERN for the very successful operation of the LHC, as well as the
support staff from our institutions without whom ATLAS could not be
operated efficiently.

We acknowledge the support of ANPCyT, Argentina; YerPhI, Armenia; ARC, Australia; BMWFW and FWF, Austria; ANAS, Azerbaijan; SSTC, Belarus; CNPq and FAPESP, Brazil; NSERC, NRC and CFI, Canada; CERN; CONICYT, Chile; CAS, MOST and NSFC, China; COLCIENCIAS, Colombia; MSMT CR, MPO CR and VSC CR, Czech Republic; DNRF and DNSRC, Denmark; IN2P3-CNRS, CEA-DRF/IRFU, France; SRNSFG, Georgia; BMBF, HGF, and MPG, Germany; GSRT, Greece; RGC, Hong Kong SAR, China; ISF and Benoziyo Center, Israel; INFN, Italy; MEXT and JSPS, Japan; CNRST, Morocco; NWO, Netherlands; RCN, Norway; MNiSW and NCN, Poland; FCT, Portugal; MNE/IFA, Romania; MES of Russia and NRC KI, Russian Federation; JINR; MESTD, Serbia; MSSR, Slovakia; ARRS and MIZ\v{S}, Slovenia; DST/NRF, South Africa; MINECO, Spain; SRC and Wallenberg Foundation, Sweden; SERI, SNSF and Cantons of Bern and Geneva, Switzerland; MOST, Taiwan; TAEK, Turkey; STFC, United Kingdom; DOE and NSF, United States of America. In addition, individual groups and members have received support from BCKDF, CANARIE, CRC and Compute Canada, Canada; COST, ERC, ERDF, Horizon 2020, and Marie Sk{\l}odowska-Curie Actions, European Union; Investissements d' Avenir Labex and Idex, ANR, France; DFG and AvH Foundation, Germany; Herakleitos, Thales and Aristeia programmes co-financed by EU-ESF and the Greek NSRF, Greece; BSF-NSF and GIF, Israel; CERCA Programme Generalitat de Catalunya, Spain; The Royal Society and Leverhulme Trust, United Kingdom. 

The crucial computing support from all WLCG partners is acknowledged gratefully, in particular from CERN, the ATLAS Tier-1 facilities at TRIUMF (Canada), NDGF (Denmark, Norway, Sweden), CC-IN2P3 (France), KIT/GridKA (Germany), INFN-CNAF (Italy), NL-T1 (Netherlands), PIC (Spain), ASGC (Taiwan), RAL (UK) and BNL (USA), the Tier-2 facilities worldwide and large non-WLCG resource providers. Major contributors of computing resources are listed in Ref.~\cite{ATL-GEN-PUB-2016-002}.
%----------------------------------------------

%-------------------------------------------------------------------------------

\printbibliography

\clearpage % ATLAS Collaboration author list
% Reference date of STDM-2018-03 is 2018-06-17
% Author list last updated on date 16-MAY-19
% Data extracted on 16-May-2019 for paper reference STDM-2018-03
% at 12:33pm
 
\begin{flushleft}
{\Large The ATLAS Collaboration}

\bigskip

M.~Aaboud$^\textrm{\scriptsize 35d}$,    
G.~Aad$^\textrm{\scriptsize 100}$,    
B.~Abbott$^\textrm{\scriptsize 127}$,    
O.~Abdinov$^\textrm{\scriptsize 13,*}$,    
B.~Abeloos$^\textrm{\scriptsize 131}$,    
D.K.~Abhayasinghe$^\textrm{\scriptsize 92}$,    
S.H.~Abidi$^\textrm{\scriptsize 166}$,    
O.S.~AbouZeid$^\textrm{\scriptsize 40}$,    
N.L.~Abraham$^\textrm{\scriptsize 155}$,    
H.~Abramowicz$^\textrm{\scriptsize 160}$
% ...
\bigskip
\\

% ...
$^{13}$Institute of Physics, Azerbaijan Academy of Sciences, Baku; Azerbaijan.\\
% ...
$^{35}$$^{(a)}$Facult\'e des Sciences Ain Chock, R\'eseau Universitaire de Physique des Hautes Energies - Universit\'e Hassan II, Casablanca;$^{(b)}$Facult\'{e} des Sciences, Universit\'{e} Ibn-Tofail, K\'{e}nitra;$^{(c)}$Facult\'e des Sciences Semlalia, Universit\'e Cadi Ayyad, LPHEA-Marrakech;$^{(d)}$Facult\'e des Sciences, Universit\'e Mohamed Premier and LPTPM, Oujda;$^{(e)}$Facult\'e des sciences, Universit\'e Mohammed V, Rabat; Morocco.\\
% ...
$^{40}$Niels Bohr Institute, University of Copenhagen, Copenhagen; Denmark.\\
% ...
$^{92}$Department of Physics, Royal Holloway University of London, Egham; United Kingdom.\\
% ...
$^{100}$CPPM, Aix-Marseille Universit\'e, CNRS/IN2P3, Marseille; France.\\
% ...
$^{127}$Homer L. Dodge Department of Physics and Astronomy, University of Oklahoma, Norman OK; United States of America.\\
% ...
$^{131}$LAL, Universit\'e Paris-Sud, CNRS/IN2P3, Universit\'e Paris-Saclay, Orsay; France.\\
% ...
$^{155}$Department of Physics and Astronomy, University of Sussex, Brighton; United Kingdom.\\
% ...
$^{160}$Raymond and Beverly Sackler School of Physics and Astronomy, Tel Aviv University, Tel Aviv; Israel.\\
% ...
$^{166}$Department of Physics, University of Toronto, Toronto ON; Canada.\\
% ...
$^{a}$ Also at Borough of Manhattan Community College, City University of New York, New York NY; United States of America.\\
$^{b}$ Also at Centre for High Performance Computing, CSIR Campus, Rosebank, Cape Town; South Africa.\\
$^{c}$ Also at CERN, Geneva; Switzerland.\\
$^{d}$ Also at CPPM, Aix-Marseille Universit\'e, CNRS/IN2P3, Marseille; France.\\
$^{e}$ Also at D\'epartement de Physique Nucl\'eaire et Corpusculaire, Universit\'e de Gen\`eve, Gen\`eve; Switzerland.\\
% ...
$^{*}$ Deceased

\end{flushleft}

% Created with Glance <Atlas.Glance@cern.ch>

\end{document}